\documentclass[prd,preprint,amsmath,nofootinbib,superscriptaddress]{revtex4}
\usepackage{graphicx,amssymb}
\usepackage{bm}
\usepackage{epsfig}

\newcommand{\beq}{\begin{equation}}
\newcommand{\eeq}{\end{equation}}
\newcommand{\bea}{\begin{eqnarray}}
\newcommand{\eea}{\end{eqnarray}}

\begin{document}

\title{Gravitational radiative corrections from effective field theory} 
\author{Walter D. Goldberger and Andreas Ross}  
\address{Department  of Physics, 
Yale University\\
New Haven CT 06520, USA}

\begin{abstract}{  
In this paper we construct an effective field theory (EFT) that describes long wavelength gravitational radiation from compact systems.    To leading order, this EFT consists of the multipole expansion, which we describe in terms of a diffeomorphism invariant point particle Lagrangian.    The EFT also systematically captures ``post-Minkowskian" corrections to the multipole expansion due to non-linear terms in general relativity.    Specifically, we compute long distance corrections from the coupling of the (mass) monopole moment to the quadrupole moment, including up to two mass insertions.    Along the way, we encounter both logarithmic short distance (UV) and long wavelength (IR) divergences.     We show that the UV divergences can be (1) absorbed into a renormalization of the multipole moments and (2) resummed via the renormalization group.    The IR singularities are shown to cancel from properly defined physical observables.    As a concrete example of the formalism, we use this EFT to reproduce a number of post-Newtonian corrections to the gravitational wave energy flux from non-relativistic binaries, including long distance effects up to 3PN ($v^6$) order.  
Our results verify that the factorization of scales proposed in the NRGR framework of Goldberger and Rothstein is consistent up to order 3PN.
}
\end{abstract}
\maketitle

\section{Introduction}

Understanding the dynamics of compact binaries within general relativity has become a problem of experimental relevance.     Gravitational wave detectors such as LIGO/VIRGO~\cite{LIGO}, or the planned LISA~\cite{LISA} are capable of probing the evolution of such systems over their entire life cycle, from the initial slow inspiral phase  to the final ringdown period after the binary constituents merge.   

In general, the two-body problem in general relativity is highly non-linear, and therefore tractable only by numerical methods.    In certain kinematic regimes, however, there is a clear separation of scales that opens up the possibility of employing analytical (perturbative) techniques to make systematic predictions. One example of this is the post-Newtonian (PN) expansion, which is possible whenever  the orbital separation $r$ is parametrically larger than the typical gravitational radius, of order $r_s = 2 G_N m$.   In that case, one can expand the Einstein equations in the velocity parameter $v^2 = r_s/r\ll 1$.    The PN expansion has been thoroughly developed by several collaborations over several decades, reaching a relative precision of $v^7$ (3.5PN) in the computations of two-body motion and of gravitational wave observables (energy-momentum and angular momentum fluxes) for non-spinning binaries.    See~\cite{PNrev} for a review and further references.    Another limit which is analytically tractable is the extreme mass ratio limit, relevant to binaries containing one supermassive black hole.   This limit can be formulated in terms of black hole perturbation theory, i.e. by finding wave solutions to the Regge-Wheeler or Teukolsky equations.   See ref.~\cite{BHpert}.

It is interesting to note that analytical control over the two-body dynamics coincides with the emergence of a wide hierarchy of scales in the system.   Field theoretic problems in which there is a separation of scales are known to admit a natural formulation in terms of effective field theories (EFTs), and in fact the quantum mechanical counterpart of the (non-relativistic) two-body problem in QED and in strong interaction physics (QCD) has been successfully addressed using EFT techniques~\cite{NREFT}.    This motivated ref.~\cite{GnR1} to recast binary dynamics in general relativity as an EFT. 

The EFT approach of~\cite{GnR1,GnR1.5} attempts to simplify both the computation of observables and the physical interpretation of terms in the perturbative expansion by treating each scale that arises in the problem one at a time, ``integrating out" short distances before moving on to the largest scales accessible to experiments.   In the two-body problem, the shortest distance scale is the size of the compact binary constituents themselves.   Integrating out this scale simply consists of writing down a Lagrangian for the center-of mass worldline of each object that consists of an infinite hierarchy of terms constructed out of the gravitational field, and suppressed by more powers of the object's radius.    Given a model for the internal structure of the compact object, the coefficients in this Lagrangian are fixed through a matching procedure discussed in more detail in ref.~\cite{GnR1}.    We stress, however, that the EFT approach is more than just employing a point particle approximation:    besides the physical radius of the binary constituents, there are several other scales in the problem,  and, as discussed in~\cite{GnR1}, properly removing these from the long distance physics necessitates a non-trivial decomposition of the gravitational field into modes with support over different kinematical regions.   

In recent years, the framework of~\cite{GnR1} has developed into a tool used to solve a number of problems in gravitational radiation and black hole dynamics within general relativity.   In particular, the formalism introduced in~\cite{GnR1} was extended to account for dissipative effects, for example absorption by black hole horizons and tidal heating in neutron stars, in refs.~\cite{GnR3,Porto2}.   Doing so necessitates the inclusion of additional degrees of freedom living on the black hole worldline.       Spin interactions were included in the formalism first in ref.~\cite{Porto1}, and used to obtain new results for various spin-spin interactions~\cite{PR1,PR2} within the PN expansion.    An EFT for the extreme mass ratio limit was constructed in~\cite{galley1, galley2} which included a treatment of gravitational radiation reaction consistent with standard results~\cite{Mino:1996nk,Quinn:1996am} at leading order.    The calculation of higher order spinless potentials within the EFT formalism, employing an extremely convenient parametrization for the metric introduced in refs.~\cite{Kol:2007rx, Kol:2007bc,Kol:2009mj}, was initiated in ref~\cite{Gilmore}, which reproduces known results at order $v^4$ and paves the way for the automatization of terms at $v^6$ and beyond.   Ref. \cite{Cannella:2008nr} computed the energy momentum tensor and \cite{Cannella:2009he} studied higher graviton vertices.   Finally, formal applications to the thermodynamic phase diagram of compactified black holes (for further references to the literature, see~\cite{KKBHrefs}) were developed within effective field theory in~\cite{GnR2.5,Kol:2007rx,smolkin}.

In this paper, we consider long wavelength gravitational radiation from compact sources within the EFT context.    At the linearized level, the relevant framework is of course simply the multipole expansion~\cite{thorne}.   In this case, the relevant expansion parameter is  $a/\lambda\ll 1$ (with $a$ the size of the source), with radiation sourced by an $\ell$-pole moment suppressed by a relative factor of $(a/\lambda)^\ell$.   As has been known for some time, in addition to the multipole expansion, there is a distinct expansion in a parameter $\eta\equiv G_N m/\lambda\ll 1$ (with $2 G_N m \le a$ the gravitational radius of the system) that arises from the non-linear nature of general relativity.   

While it is straightforward to implement the naive multipole expansion (see for instance the textbook~\cite{maggiore}), the $\eta$-expansion has a much richer structure~\cite{burke,postM1,tail1,postM2}. Long wavelength (IR) logarithmic singularities arise already at order $\eta^1$, and at $\eta^2$ there are both IR, as well as short distance (UV) logarithmic divergences\footnote{The conventional nomenclature for these effects is the ``tail" of the gravitational waves~\cite{tail}.   These have been computed in ref.~\cite{tail1}  (at order $\eta^1$), and at order $\eta^2$ in ref.~\cite{Blanchet:1997jj}.}.     In sec.~\ref{sec:postM}, we set up a systematic expansion in powers of $\eta$ in the EFT language of~\cite{GnR1}.    

In order to accomplish this, we begin in sec.~\ref{sec:setup} by assuming that the short distance (multipole) scale $a$ has been integrated out.   This results in an effective action consisting of a point particle with worldline localized sources (corresponding to the moments of the system), coupled to gravity.   This effective Lagrangian is manifestly invariant under both diffeomorphisms and worldline reparametrizations.    In sec.~\ref{sec:obs} we discuss what observables can be computed using this Lagrangian.    

Given these preliminaries, we set out to explicitly compute corrections in the post-Minkowskian parameter $\eta$, working up to order $\eta^2$ (and therefore order 3PN for non-relativistic systems).    We find that there are IR divergences at the level of graviton emission amplitudes, essentially due to the long range of the Newtonian gravitational potential.   However,  we show that these cancel explicitly from observable quantities, at least to order $\eta^2$.    In sec.~\ref{sec:IRsum}, we consider IR effects at higher orders in perturbation theory.   Using a line of analysis first developed in~\cite{Weinberg:1965nx}, we estimate that the leading IR singularity at order $\eta^n$ scales like the $n$-th power of a divergent logarithm.  Consequently, we find that summing the  leading logarithmic IR divergences to all orders results in an overall phase factor in the amplitude, therefore canceling from physical observables.   This gives partial indication that the formalism is free of long wavelength singularities at all orders in the $\eta$ expansion.

In addition to IR logarithms, at order $\eta^2$ there are UV divergent logarithms.    Unlike the IR effects, the UV singularities are true singularities of the long wavelength EFT, which must be dealt with by renormalizing the coupling constants of the theory (i.e., the moments).    The procedure for doing so is described in sec.~\ref{sec:RG}.  There we show how the introduction of renormalized moments renders finite the graviton amplitudes to order $\eta^2$.    (We focus only on the quadrupole moment, although the same procedure applies to the higher moments as well.).    The renormalized moments exhibit subtraction scale dependence governed by a non-trivial renormalization group (RG) equation.   In principle, this RG equation generates and resums the entire series of ``leading UV logarithms" at any given order in the expansion parameter.  Although in practical situations the logarithms do not seem to become large enough to necessitate this resummation, the RG scaling is theoretically interesting as it gives quantitative constraints on the form of the perturbative expansion.

The analysis of sec.~\ref{sec:postM} is universal, and applies to long wavelength radiation from any localized system.   In order to make definite predictions, one must specify the multipole moments of the EFT, by  \emph{matching} to a more complete theory that describes the short distance gravitational dynamics of the system.  Matching is standard in EFTs, but we outline how the procedure is carried out in the present context in sec.~\ref{sec:mult}.    For illustration purposes, we apply our general framework to post-Newtonian systems in sec.~\ref{sec:PN}.   There we use the results of sec.~\ref{sec:mult} together with the NRGR formalism~\cite{GnR1}, an EFT for gravitating non-relativistic systems, to compute the multipole moments needed for the 1PN corrections to the energy flux.   Together with sec.~\ref{sec:postM} this allows us to reproduce PN corrections for spinless systems at the orders $v^2$, $v^3$, and $v^5$ beyond leading order.   In addition we compute the non-analytic terms $v^6\ln v$ at 3PN.   Finally, we present our conclusions in sec.~\ref{sec:conclusions}.

\section{EFT setup}
\label{sec:setup}

Consider an arbitrary compact source, e.g, a binary system, emitting gravitational radiation.    In this paper we focus on the part of the wave spectrum with wavelength much larger than the characteristic size of the source.    It is then natural to describe this radiation in terms of source multipole moments.   Formally, this means that one can systematically decompose the motion of the system in terms of a central worldline $x^\mu(\tau)$ that traces the source's path through spacetime, as well as a set of moments that describe the internal dynamics.   These moments are a set of dynamical degrees of freedom localized on this worldline, labeled by their transformation properties under rotations in the rest frame and by transformations under parity.

In equations, this information is encoded in the Lagrangian~\cite{GnR3}
\begin{eqnarray}
\label{eq:lag}
\nonumber
S &=&  - m\int d\tau - \frac{1}{2}\int dx^\mu  L_{ab}\, \omega^{ab}_\mu(\tau)+{1\over 2}  \sum_{n=0}^\infty \int d\tau c^{(I)}_n I^{a b a_1\ldots a_n}(\tau) \nabla_{a_1}\cdots\nabla_{a_{n}} E_{a b}(x)\\
& & +{1\over 2} \sum_{n=0}^\infty \int d\tau c^{(J)}_n J^{a b a_1\ldots a_n}(\tau) \nabla_{a_1}\cdots\nabla_{a_{n}} B_{a b}(x).
\end{eqnarray}      
Here, $d\tau=\sqrt{g_{\mu\nu} dx^\mu dx^\nu}$ is proper time along the central worldline.     We have chosen a local Lorentz frame at each point $x^\mu(\tau)$ in such a way that $e^\mu_0 = v^\mu = dx^\mu/d\tau$ is the four-velocity, and $e^\mu_a(\tau)$, $a=1,2,3$ comprises a spatial frame whose rotation encodes spin dynamics\footnote{There is considerable freedom in the choice of local Lorentz frame, and therefore in the choice of variables to describe spin.    See~\cite{Porto1,PR1,PR2} for further discussions.}.    Our choice of frame therefore satisfies 
\begin{equation}
g^{\mu\nu} = e^\mu_0 e^\nu_0 - \delta^{ab} e^\mu_a e^\nu_b.
\end{equation}
The components  $\omega^{ab}_\mu$ of the spin connection\footnote{$\omega^{ab}_\mu$ is related to the usual spin connection $\omega^{AB}_\mu$ for the metric vierbein $E^{A=0,\dots,3}_\mu(x)$, $g_{\mu\nu}(x) = \eta_{AB} {E^A}_\mu {E^B}_\nu,$ by the change of basis ${\Lambda^a}_A(\tau)\equiv e^a_\mu(\tau) {E^\mu}_A(x(\tau))$.} couple to the total angular momentum $L^{ab}(\tau)=-L^{ba}(\tau)$ of the compact source.     Finally,  the $SO(3)$ tensors $I^{a_1\ldots a_\ell}$, $J^{a_1,\ldots a_\ell}$ ($\ell\geq 2$), taken respectively to be of electric and magnetic parity, and symmetric and traceless with respect to the Euclidean metric $\delta_{ab}$, define the full set of multipole moments of the compact object.   For later convenience, we have chosen the normalization constants in the definitions to be
\begin{equation}
c^{(I)}_0 = 1, \ \  c^{(J)}_0 = - {4\over 3} , \ \  c^{(I)}_1 = {1\over 3}.
\end{equation}    
We will not use any higher moments for the calculations in this paper.     The moments $I^{a_1\ldots a_\ell}$, $J^{a_1,\ldots a_\ell}$ ($\ell\geq 2$) serve as sources of gravitational radiation and therefore couple to the electric and magnetic parity curvature tensors, defined in terms of the Weyl tensor as
\bea
E_{ab} &=& C_{\mu \alpha \nu \beta}v^{\mu} e^\alpha_a v^\nu e^\beta_b  \label{eq:WeylErel}, \\
B_{ab} &=& - \frac{1}{2}\epsilon_{acd}  e^c_\mu e^d_\nu {C^{\mu\nu}}_{\lambda\sigma} e^\lambda _bv^\sigma \label{eq:WeylBrel},
\eea
and for $\ell>2$  their covariant gradients, obtained by operating with $\nabla_a = e^\mu_a \nabla_\mu$, for $\ell>2$.    Note that  $E_{ab}=E_{ba}$, $\delta^{ab} E_{ab}=0,$ and likewise for $B_{ab}$.   Finally, note that in Eq.~(\ref{eq:lag}) we have only kept terms that are linear in the curvature tensor.   Terms quadratic and higher, such as $\int d\tau E_{ab} E^{ab}$, $\int d\tau B_{ab} B^{ab}$, were first introduced in~\cite{GnR1,GnR1.5} and encode the tidal response of the localized system to external gravitational fields.

By construction, Eq.~(\ref{eq:lag}) provides a fully diffeomorphism invariant description of a compact source, in the limit of long wavelength radiation. Together with the Einstein term $S_{EH}=-2 m_{Pl}^2 \int d^4 x \sqrt{g} R$, with $m_{Pl}^2\equiv 1/(32\pi G_N)$, Eq.~(\ref{eq:lag}) can be used as a starting point for a systematic long wavelength expansion, both in the multipole expansion parameter $a/\lambda\ll 1$ and the ``post-Minkowskian" (borrowing the terminology of ~\cite{postM1}) parameter $\eta = G_N M/\lambda\ll 1$ that controls non-linear corrections.  

In order to compute these corrections systematically, one must specify (as in any other EFT) a power counting scheme for bookkeeping relative sizes of terms.   Since we are interested in radiation of wavelength characterized by a scale $\lambda$, we assume that derivatives acting on the gravitational field scale as $\partial_\mu\sim \lambda^{-1}$.     Hence, we take $x^\mu\sim\lambda$, and by demanding that the kinetic term for  $h_{\mu\nu}/m_{Pl}=g_{\mu\nu}-\eta_{\mu\nu}$ be leading order, $h_{\mu\nu}\sim \lambda^{-1}$.   The $\ell$-th multipole moment is expected to scale as $m a^\ell$, and thus multipole couplings in Eq.~(\ref{eq:lag}) are suppressed\footnote{In practice, for instance in applications to PN systems, the $\ell$-th multipole scales not as a definite power, but rather as a series $m a^\ell\left[1+{\cal O}(a/\lambda) + {\cal O}(a/\lambda)^2 + \cdots\right]$.   Strictly speaking, to have manifest power counting one would have to introduce a separate coupling in Eq.~(\ref{eq:lag}) for each term in this series.    This would be extremely awkward to write out explicitly, so we will stick with the standard conventions in the general relativity literature of classifying moments by their $SO(3)$ transformation properties rather than by their scaling with respect to the expansion parameter.} by powers of $a/\lambda\ll1$.    With these rules, it is possible to determine the scaling in powers of $\eta$ or $a/\lambda$ of any term in the action.   For instance at a fixed order in $a/\lambda$, the post-Minkowskian terms are generated by Feynman graphs with increasing numbers of insertions of the mass monopole term $-m\int d\tau\sim m\lambda$ and $n$-graviton vertices, which scale as $(m_{Pl}\lambda)^{2-n}$.

The rest of this paper is devoted to explicitly computing corrections to low frequency gravitational wave emission, both in powers of $\eta$ and in powers of the multipole expansion parameter.   In the next section we give a general discussion of how such observables are computed in practice, given a prescribed set of moments $I^{a_1\ldots a_\ell}(\tau),$ $J^{a_1\ldots a_\ell}(\tau)$.     In sec.~\ref{sec:postM} we use this Lagrangian to compute corrections to gravitational radiation suppressed by powers of $\eta,$ working up to order $\eta^2$.   Finally in sec.~\ref{sec:mult} we discuss how to determine the moments $I^{a_1\ldots a_\ell},$ $J^{a_1\ldots a_\ell}$ for a localized gravitational system.   We apply these results to post-Newtonian systems in sec.~\ref{sec:PN}.

\subsection{Calculating observables}
\label{sec:obs}

As in our previous work, we find it convenient to set up diagrammatic rules for the computation of observables associated with the emission of gravitational radiation (e.g., energy, momentum, and angular momentum flux).              In this paper, we will restrict ourselves to time averaged observables (a field theoretic formulation of instantaneous quantities is given in~\cite{galley2}.)    A natural way to set up the diagrammatics is to formulate all observables in terms of the matrix element for the emission of a single graviton from the compact source described by Eq.~(\ref{eq:lag}),
\begin{equation}
i \mathcal A_h({\bf k}) =  \parbox{21.5mm}{\includegraphics{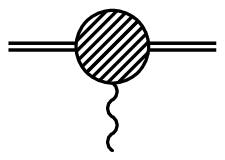}}.
\end{equation}
Here ${\cal A}_h({\bf k})$ denotes the probability amplitude to emit a graviton of momentum ${\bf k}$ and definite helicity $h=\pm 2$ (as measured in a nearly Lorentz frame infinitely far from the source).   This is given by Feynman diagrams with one on-shell external graviton (internal vertices and propagators are obtained from the source terms in Eq.~(\ref{eq:lag}) plus the Einstein-Hilbert action $S_{EH}$).

In terms of ${\cal A}_h({\bf k})$ one can compute a polarized graviton emission rate as
\begin{equation}
d\Gamma_h({\bf k}) = {1\over T} {d^3 {\bf k}\over (2\pi)^3 2|{\bf k}|} |{\cal A}_h({\bf k})|^2,
\end{equation} 
where $T\rightarrow\infty$ represents total integration in the detector, and drops out of time averaged quantities.   Moments of the differential rate $d\Gamma_h({\bf k})$ give rise to physical observables.   For example, the (polarized) total rate of radiated linear four-momentum is given by
\begin{equation}
\left.{\dot P^\mu}\right|_{h=\pm 2} = \int k^\mu d\Gamma_h({\bf k}),
\end{equation}
where $k^\mu=(|{\bf k}|,{\bf k})$ is the four-momentum of the emitted graviton\footnote{Strictly speaking, there should be a cut in the integration over graviton frequency at a value $\omega_*\sim 1/a$ where the multipole expansion begins to breaks down.}.    It is also possible to compute the rate of angular momentum loss by the system.   In terms of the helicity amplitudes ${\cal A}_h({\bf k})$, it is 
\begin{equation}
{\dot {\bf J}} = \sum_h \int  h \hspace*{0.8pt} {\bf n} \hspace*{0.8pt} d\Gamma_{h}({\bf k}),
\end{equation}
where ${\bf n} = {\bf k}/|{\bf k}|$ is the direction of the emitted graviton and we sum over the physical helicities $h = \pm 2$.

As an example, ignoring post-Minkowskian corrections, Eq.~(\ref{eq:lag}) gives for the one-graviton matrix element,
\begin{align}
i \mathcal A_h({\bf k}) & =  \parbox{23mm}{\includegraphics{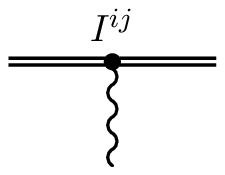}} + \hspace*{2pt} \parbox{23mm}{\includegraphics{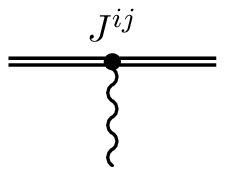}} + \hspace*{2pt} \parbox{23mm}{\includegraphics{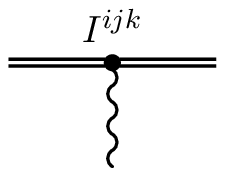}} + \cdots \notag \\
                 & = {i\over  4m_{Pl}} \epsilon^*_{ij}({\bf k}, h)  \left[{\bf k}^2 I^{ij}(k) + {4\over 3} |{\bf k}| \, {\bf k}^l \epsilon^{ikl}J^{jk}(k) -{i\over 3} {\bf k}^2 {\bf k}^l I^{ijl}(k) + \cdots \right] \label{eq:ampliIIJ},
\end{align}
where we work with physical on-shell graviton degrees of freedom with polarization tensor $\epsilon_{ij}({\bf k},h)$ satisfying the usual conditions ${\bf k}^i \epsilon_{ij}({\bf k},h)= \epsilon_{ii}({\bf k},h)=0$.   Here we have assumed for simplicity that the source is at rest, and take the frame $\{e^0,e^a\}$ to align with the Lorentz frame at infinity (in which case the rotation of the system is reflected in the time dependence of the moments).    Then using the standard result for the sum over graviton polarizations
\begin{eqnarray}
\nonumber
\sum_h \epsilon_{ij}({\bf k},h) \epsilon^{*}_{rs}({\bf k}, h)&=& {1\over 2}\left[\delta_{ir}\delta_{js}+\delta_{is}\delta_{jr}-\delta_{ij}\delta_{rs} +{1\over  {\bf k}^2}\left(\delta_{ij} {\bf k}_r {\bf k}_s +\delta_{rs} {\bf k}_i {\bf k}_j\right)\right. \\
& & \left. {}-{1\over {\bf k}^2}\left(\delta_{ir} {\bf k}_j {\bf k}_s + \delta_{is} {\bf k}_j {\bf k}_r + \delta_{jr} {\bf k}_i {\bf k}_s + \delta_{js} {\bf k}_i {\bf k}_r\right) + {1\over {\bf k}^4} {\bf k}_i {\bf k}_j {\bf k}_r {\bf k}_s\right], \ \  
\end{eqnarray}
one finds  for the total radiated power 
\begin{equation}
{\dot P}^0 =   {G_N\over \pi T} \int^\infty_0 dk  \left[ {k^6\over 5} \left|I^{ij}(k)\right|^2 + {16\over 45} k^6\left|J^{ij}(k)\right|^2+ {k^8\over 189}\left|I^{ijk}(k)\right|^2 + \cdots\right] \label{eq:PowerIJImom}.
\end{equation}
Transforming to the time domain, one then gets the standard result~\cite{maggiore}
\begin{equation}
{\dot P^0}  =   {G_N\over 5} \left< \left({d^3\over  dt^3} I^{ij}(t)\right)^2 \right>+ {16 G_N\over 45} \left< \left({d^3\over  dt^3} J^{ij}(t)\right)^2 \right>+  {G_N \over 189} \left< \left({d^4\over  dt^4} I^{ijk}(t)\right)^2 \right> + \cdots\label{eq:PowerIJIcoo},
\end{equation}
where the time average is $\left< \cdots \right>={1\over T}\int_0^T \cdots,$ as $T\rightarrow\infty$. 

\section{Post-Minkowskian corrections}
\label{sec:postM}

At the linearized level, long wavelength radiation from a localized source is described by Eq.~(\ref{eq:ampliIIJ}).    We now show how to compute corrections in the EFT  arising from graviton self-interactions.   Not only are these corrections relevant for gravitational wave phenomenology, but they also introduce logarithmic singularities at both long and short distances.   In this section, we discuss the physical origin of these divergences and explain how they are systematically removed from observable quantities.

\subsection{Corrections at order $\eta^1$}

\begin{figure}[!t]
\centerline{{\includegraphics{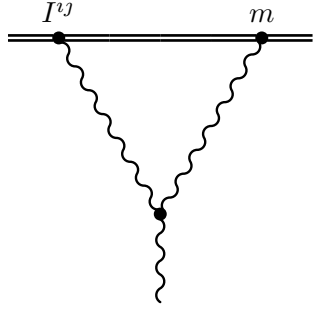}}}
\caption[1]{Leading post-Minkowskian correction.} \label{fig:tail1}
\vskip 0.5cm
\end{figure}

Long distance (infrared) logarithmic singularities appear already at order $\eta^1$, from the graph in Fig.~\ref{fig:tail1}.   This graph represents the interaction of outgoing gravitational radiation (sourced, for example, by the system's electric quadrupole) with the Newtonian potential generated by the total mass of the system.   
Using the Feynman rules derived in~\cite{GnR1},  we find that after projecting onto transverse traceless parts and reducing tensor integrals to scalars via standard methods (see e.g.,~\cite{smirnov}), the graph is proportional to 
\begin{equation}
i{\cal A}_{\eta^1}= i {\cal A}_{\eta^0}\left(G_N m {\bf k}^2\right)\times I({\bf k}),
\end{equation}
where  
\begin{equation}
i \mathcal A_{\eta^0} = \frac{i {\bf k}^2}{4 m_{Pl}} \, \epsilon^*_{ij}({\bf k},h)\, I^{ij} (|{\bf k}|)
\label{eq:AquadLO},
\end{equation}
is the leading order quadrupole radiation amplitude, and the function $I({\bf k})$ is a linear combination of integrals of the form (with $d$ denoting the dimension of spacetime)
\begin{equation}
\left({1\over {\bf k}^2}\right)^n  \int {d^{d-1} {\bf q}\over (2\pi)^{d-1}} \left({1\over {\bf q}^2}\right)^{1-n} {1\over {\bf k}^2 - ({\bf k} + {\bf q})^2 + i\epsilon},
\end{equation}
with $\epsilon\rightarrow 0^+$.

The general structure of $I({\bf k})$ is not difficult to understand.   Focus on the term with $n=0$.   Then the factor of $1/{\bf q}^2$ is the Fourier transform of the Newtonian potential of the source, while the factor $1/{\bf k}^2 - ({\bf k} + {\bf q})^2$ is the propagator for a graviton of energy $|{\bf k}|$ emitted from the quadrupole vertex.   Note that in $d=4$ dimensions, the integrand for $n=0$ has the long distance (${\bf q}\rightarrow 0$) behavior
\begin{equation}
\int {d^3 {\bf q}\over (2\pi)^3}  {1\over {\bf q}^2} {1\over {\bf k}\cdot {\bf q}},
\end{equation}
and is therefore \emph{infrared} logarithmically divergent.   This IR singularity is physically due to the interaction between the nearly on-shell emitted graviton and the long range $1/r$ potential of the source over an infinitely large amount of time.   It is entirely analogous to the divergent phase factor that appears in quantum mechanical Coulomb scattering.   

In order to regularize this divergence, we use dimensional regularization, keeping the spacetime dimension $d$ arbitrary in the calculations and analytically continuing to a neighborhood of $d=4$ in the complex $d$-plane.   With the aid of the table of integrals in appendix~\ref{app:ints}, we find for the ${\cal O}(\eta^1)$ amplitude
\begin{equation}
\label{eq:eta1amp}
i {\cal A}_{\eta^1} = i {\cal A}_{\eta^0} \times \left(i G_N m |{\bf k}|\right) \left[- {({\bf k}^2 + i\epsilon)\over \pi \mu^2} e^{\gamma_E}\right]^{(d-4)/2}\times \left[{2\over d-4} - {11\over 6} + (d-4)\left({\pi^2\over 8} + {203\over 72}\right)+\cdots\right],
\end{equation}
where terms of order $(d-4)^2$ and higher have been dropped.  

 The scale $\mu$ is an arbitrary subtraction scale put in by hand to make the expression dimensionally correct as $d\rightarrow 4$.   Formally, it arises by introducing a ``renormalized" Newton's constant,
\begin{equation}
-2 {m^B_{Pl}}^2 \int d^4 x \sqrt{g} R = - 2 Z^{-1}_G (\mu) m_{Pl}^2 \mu^{d-4} \int d^d x \sqrt{g} R,
\end{equation}
or $G^B_N= Z_G(\mu) \mu^{4-d} G_N$, with renormalization factor $Z_G=1+{\cal O}(\hbar)$.   The $\mu$ independence of the bare theory then implies a (classically trivial) renormalization group equation for the renormalized Newton's constant,
\begin{equation}
\label{eq:GRG}
\mu {d\over d\mu}  G_N =  (d-4) G_N,
\end{equation}
which will play a role later in Sec.~\ref{sec:RG}.    From now on, only the renormalized $G_N$ will appear in our calculations.

At order $\eta^1$, the $\mu$ dependence explicitly drops out of physical quantities, however.   This follows since at this order,  the squared modulus of the emission amplitude is
\begin{equation}
\label{eq:amp2eta1}
\left|{{\cal A}\over {\cal A}_{\eta^0}}\right|^2 = 1 + 2 \mbox{Re} {{\cal A}_{\eta^1}\over {\cal A}_{\eta^0}} + {\cal O}(\eta^2),
\end{equation}
and from Eq.~(\ref{eq:eta1amp}), 
\begin{eqnarray}
\nonumber
\mbox{Re} {{\cal A}_{\eta^1}\over {\cal A}_{\eta^0}} &=&  (G_N m |{\bf k}|) \left|{{\bf k}^2\over \pi \mu^2} e^{\gamma_E}\right|^{(d-4)/2} \sin \pi \left({d-4\over 2}\right) \left[{2\over d-4} - {11\over 6} + (d-4)\left({\pi^2\over 8} + {203\over 72}\right)+\cdots\right],
\end{eqnarray}
Hence as $d\rightarrow 4$ along the real axis,
\begin{equation}
\label{eq:eta1tail}
\left|{{\cal A}\over {\cal A}_{\eta^0}}\right|^2 = 1 + 2\pi G_N m |{\bf k}|.
\end{equation}
Although we have only presented this result for quadrupole emission, it is in fact universal, and applies to radiation from any multipole moment\footnote{This follows from the fact that this term is determined by the coefficient of a logarithmic IR singularity. We have, in fact, explicitly checked that Eq.~(\ref{eq:eta1tail}) also holds for magnetic quadrupole and electric octupole radiation.}.   Note that, upon squaring, Eq.~(\ref{eq:eta1amp}) also produces infrared single and double poles in $d-4$ at order $\eta^2$.    It is an important check of the formalism that these poles cancel from physical quantities, as we now show.

\subsection{Effects at order $\eta^2$}
\label{sec:eta2}

\begin{figure}[!t]
\centerline{{\includegraphics{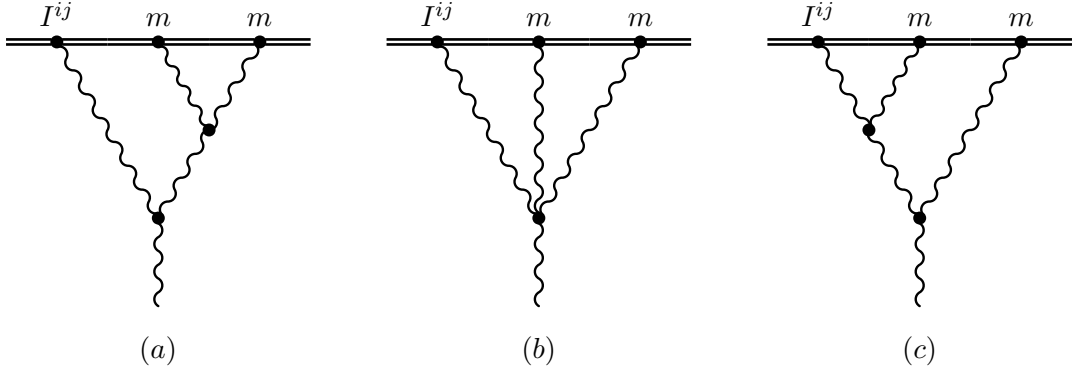}}}
\caption[1]{Second order post-Minkowskian corrections.} \label{fig:tail2}
\vskip 0.5cm
\end{figure}

The order $\eta^2$ corrections to quadrupole emission are shown in Fig.~\ref{fig:tail2}.   In addition to infrared divergences similar in nature to those encountered at order $\eta^1$, a new feature at this order is the appearance of logarithmic short distance (UV) singularities, whose physical origin and resolution will be discussed below.   

Operationally, the presence of UV divergences can be seen by examining the structure of the graphs in Fig.~\ref{fig:tail2}.   Take for example Fig.~\ref{fig:tail2}(b), which is proportional to the integral
\begin{equation}
\label{eq:tube}
\int {d^{d-1}{\bf q}\over (2\pi)^{d-1}} {1\over {\bf k}^2 - ({\bf k}+{\bf q})^2} \int {d^{d-1}{\bf p}\over (2\pi)^{d-1}} {1\over {\bf p}^2} {1\over ({\bf p}+{\bf q})^2} 
\end{equation}
where the tensor structure of has been suppressed for clarity.   By dimensional analysis, the integral over ${\bf p}$ scales as $(1/{\bf q}^2)^{(5-d)/2}$, and in coordinate space reflects the interaction of the emitted graviton with the $(G_N m/r)^2$ relativistic potential (in $d=4$) of the source.   (This is particularly clear in Fig.~\ref{fig:tail2}(a), which contains a subgraph with two $m$ insertions and one internal three-graviton vertex.   This is precisely the Feynman diagram corresponding to the $1/r^2$ potential.).      From the scaling of the ${\bf p}$ integral in Eq.~(\ref{eq:tube}), we deduce that the short distance (${\bf q}\rightarrow\infty$) behavior of Fig.~\ref{fig:tail2}(b) is, in $d=4$,
\begin{equation}
\int {d^3{\bf q}\over (2\pi)^3} {1\over |{\bf q}|} {1\over {\bf k}^2 - ({\bf k}+{\bf q})^2}\rightarrow \int {d^3{\bf q}\over (2\pi)^3} {1\over |{\bf q}|^3}.
\end{equation}
Fig.~\ref{fig:tail2}(b) is therefore logarithmically \emph{ultraviolet} divergent.    On the other hand, the integrand behaves as $1/(|{\bf q}| {\bf q}\cdot {\bf k})$ in the infrared limit ${\bf q}\rightarrow 0$, so  Fig.~\ref{fig:tail2}(b) is infrared finite by power counting.   A similar estimate reveals identical UV and IR behavior for the graph in  Fig.~\ref{fig:tail2}(a).   

Performing a chain of reductions to scalar integrals, projecting onto transverse traceless polarizations, and employing a set of standard Feynman integrals collected in appendix~\ref{app:ints}, we find
\begin{equation}
{{\cal A}^{(a)}_{\eta^2}\over {\cal A}_{\eta^0}}  =  {89\over 63} \left (G_N m |{\bf k}|\right)^2 \left[- {({\bf k}^2 + i\epsilon)\over \pi \mu^2} e^{\gamma_E}\right]^{(d-4)}\times\left[{1\over d-4} -{14989\over 18690}\right],
\end{equation}
and
\begin{equation}
{{\cal A}^{(b)}_{\eta^2}\over {\cal A}_{\eta^0}}  =  {8\over 9} \left (G_N m |{\bf k}|\right)^2 \left[- {({\bf k}^2 + i\epsilon)\over \pi \mu^2} e^{\gamma_E}\right]^{(d-4)} \times\left[{1\over d-4} -{157\over 15}\right].
\end{equation}
In these expressions, poles in $d-4$ denote the purely ultraviolet divergences discussed above.  

Evaluation of the graph in Fig.~\ref{fig:tail2}(c) requires more care, as it contains both UV and IR singularities.  Suppressing the tensor structure from the three-graviton vertices, the propagator structure of  Fig.~\ref{fig:tail2}(c) is
\begin{equation}
I^{(c)}_{\eta^2}({\bf k}) = \int {d^{d-1}{\bf q}\over (2\pi)^{d-1}}  {d^{d-1}{\bf p}\over (2\pi)^{d-1}}  {1\over {\bf q}^2}  {1\over {\bf k}^2 - ({\bf k}+{\bf q})^2+i\epsilon} {1\over {\bf p}^2}  {1\over {\bf k}^2 - ({\bf k}+{\bf p}+{\bf q})^2+i\epsilon},   
\end{equation}
which represents the amplitude for the emitted graviton to interact twice with the source's Newton potential.   Infrared divergences come from the region of integration ${\bf q},{\bf p}\rightarrow 0$.   In this limit, the ${\bf q},{\bf p}$ integrals factorize, and thus
\begin{equation}
I^{(c)}_{\eta^2}({\bf k})\rightarrow  \left[\int {d^3 {\bf q}\over (2\pi)^3}  {1\over {\bf q}^2} {1\over {\bf k}\cdot {\bf q}}\right]^2.
\end{equation}
In $d$ dimensions, this produces an infrared double pole $1/(d-4)^2$ which, as we will see explicitly below, exactly cancels the infrared double pole from the square of the amplitude ${\cal A}_{\eta^1}$ computed above.   

In order to power count the UV divergences of diagram Fig.~\ref{fig:tail2}(c) we need to be more careful about  the dependence of the numerator on the momenta. The diagram contains two insertions of graviton self-interaction vertices, each scaling as two derivatives.   This  introduces in total four powers of momenta in the numerator.   Of all the terms in the numerator, we focus on a term in the amplitude proportional to $\mathbf k^2 \mathbf q^2 $. For this component,  the behavior of the integral in the UV region is now
\begin{equation}
\tilde I^{(c)}_{\eta^2}({\bf k})\rightarrow \int {d^{d-1}{\bf q}\over (2\pi)^{d-1}}  {1\over {\bf q}^2} \int {d^{d-1}{\bf p}\over (2\pi)^{d-1}}   {1\over {\bf p}^2} {1\over ({\bf p}+{\bf q})^2}.
\end{equation}
Given the  $|{\bf q}|^{d-5}$ scaling of the ${\bf p}$ integral, for $d=4$ the ${\bf q},{\bf p}\rightarrow\infty$ behavior is UV logarithmically divergent,
\begin{equation}
\tilde I^{(c)}_{\eta^2}({\bf k})\rightarrow \int {d^3{\bf q}\over (2\pi)^3}  {1\over |{\bf q}|^3}.
\end{equation}

The exact calculation of  Fig.~\ref{fig:tail2}(c) (or ``mother of all tails") is somewhat involved.   We have found it useful to apply the Mellin-Barnes method for calculating multi-loop Feynman integrals, as reviewed for instance in Ref.~\cite{smirnov}.   Leaving the details of the calculation for appendix~\ref{app:ints}, our result for Fig.~\ref{fig:tail2}(c)  is
\begin{equation}
{{\cal A}^{(c)}_{\eta^2}\over {\cal A}_{\eta^0}}  =   \left (G_N m |{\bf k}|\right)^2 \left[- {({\bf k}^2 + i\epsilon)\over \pi \mu^2} e^{\gamma_E}\right]^{(d-4)} \times\left[-{2\over (d-4)^2 } +{109\over 315}{1\over d-4} -{7\pi^2 \over 12} + {1364777\over 132300}\right],
\end{equation}
where, as discussed above, the double pole in $d-4$ indicates an IR divergence.   On the other hand, the single pole contains both UV and IR logarithmic singularities.   It is useful to explicitly isolate the UV pole.   This is done in appendix~\ref{app:poles}, with the result
\begin{equation}
\left. {{\cal A}^{(c)}_{\eta^2}\over {\cal A}_{\eta^0}} \right|_{UV}    =    \left (G_N m |{\bf k}|\right)^2 \left[- {({\bf k}^2 + i\epsilon)\over \pi \mu^2} e^{\gamma_E}\right]^{(d-4)} \times \left[-{1046\over 315} {1\over d-4}\right] ,
\end{equation}
so we can write
\begin{eqnarray}
\label{eq:eta2ampc}
\nonumber
{{\cal A}^{(c)}_{\eta^2}\over {\cal A}_{\eta^0}}  &=&   \left (G_N m |{\bf k}|\right)^2 \left[- {({\bf k}^2 + i\epsilon)\over \pi \mu^2} e^{\gamma_E}\right]^{(d-4)} \times\left[-{2\over (d-4)_{IR}^2 } +{11\over 3}{1\over (d-4)_{IR}} - {1046\over 315}{1\over (d-4)_{UV}}\right.\\
& & {} \left. -{7\pi^2 \over 12} + {1364777\over 132300}\right].
\end{eqnarray}
The total amplitude at order $\eta^2$ is then
\begin{eqnarray}
\label{eq:ampeta2}
\nonumber
{{\cal A}_{\eta^2}\over {\cal A}_{\eta^0}}  &=&   \left (G_N m |{\bf k}|\right)^2 \left[- {({\bf k}^2 + i\epsilon)\over \pi \mu^2} e^{\gamma_E}\right]^{(d-4)} \times\left[-{2\over (d-4)_{IR}^2 } + {11\over 3}{1\over (d-4)_{IR}} - {107\over 105}{1\over (d-4)_{UV}}\right.\\
& &  \left. -{7\pi^2 \over 12} - {1777\over 14700}\right].
\end{eqnarray}
To compute physical quantities at order $\eta^2$, one needs the full modulus squared, including terms from the square of the amplitude ${\cal A}_{\eta^1}$ calculated in the previous section.  The full set of terms at ${\cal O}(\eta^2)$ is
\begin{equation}
\left|{{\cal A}\over {\cal A}_{\eta^0}}\right|_{\eta^2} ^2 =  2 \mbox{Re} {{\cal A}_{\eta^2}\over {\cal A}_{\eta^0}} + \left|{{\cal A}_{\eta^1}\over {\cal A}_{\eta^0}}\right|^2,
\end{equation}
and from Eq.~(\ref{eq:eta1amp}), Eq.~(\ref{eq:eta2ampc}),
\begin{equation}
\left|{{\cal A}_{\eta^1}\over {\cal A}_{\eta^0}}\right|^2 =  \left (G_N m |{\bf k}|\right)^2 \left|{{\bf k}^2\over \pi \mu^2} e^{\gamma_E}\right|^{(d-4)}\left[ {4\over (d-4)_{IR}^2} -{22\over 3} {1\over (d-4)_{IR}} +{\pi^2\over 2} +{527\over 36}\right],
\end{equation}
\begin{equation}
2 \mbox{Re} {{\cal A}_{\eta^1}\over {\cal A}_{\eta^0}}  = \left (G_N m |{\bf k}|\right)^2 \left|{{\bf k}^2\over \pi \mu^2} e^{\gamma_E}\right|^{(d-4)}\left[ -{4\over (d-4)_{IR}^2}+{22\over 3} {1\over (d-4)_{IR}} -{214\over 105} {1\over (d-4)_{UV}} + {5\pi^2\over 6} -{1777\over 7350}\right].
\end{equation}
Comparing the two expressions, we see the explicit cancellation of infrared divergences.      Our final result is
\begin{eqnarray}
\label{eq:amp2eta2}
\nonumber
\left|{{\cal A}\over {\cal A}_{\eta^0}}\right|^2  &=& 1 + 2 \pi\left (G_N m |{\bf k}|\right)  + \left (G_N m |{\bf k}|\right)^2 \left[ -{214\over 105}\left( {1\over (d-4)_{UV}} +\gamma_E + \ln {{\bf k}^2\over \pi\mu^2}\right)\right. \\ 
& & {}\left. + {4\pi^2\over 3} + {634913\over 44100}\right]  + {\cal O}(\eta^3).
\end{eqnarray}

Although the infrared divergences have disappeared to ${\cal O}(\eta^2)$, the ultraviolet divergence remains.   In the PN context, the presence of such logarithmic singularities was first pointed out in ref.~\cite{Anderson:1982fk}.   See also~\cite{Blanchet:1997jj}.    This UV divergence represents a true short distance singularity of the EFT, and requires renormalization of the theory as we now discuss.

\subsection{Renormalization}
\label{sec:RG}

It is straightforward to identify a suitable counterterm that cancels the UV pole in Eq.~(\ref{eq:ampeta2}).    From the form of the leading order quadrupole amplitude, we can interpret the frequency domain quadrupole moment as a (frequency dependent) bare coupling constant.   Then the explicit regulator dependence of the amplitude cancels that of the bare coupling, leaving behind a finite remainder.   

To make this more explicit, we  begin by introducing a renormalized quadrupole moment,     
\begin{equation}
\label{eq:Irdef}
I^B _{ij}(\omega) = Z(\omega,\mu) I^R_{ij}(\omega, \mu)
\end{equation}
where we denote the wave frequency $\omega = |\mathbf k|$. Note that in order to keep the bare moment $I^B _{ij}$ scale independent, we must introduce $\mu$ dependence in the renormalized moment $I^R_{ij}$.    If the renormalization constant $Z(\omega,\mu)$ is adjusted as
\begin{equation}
\label{eq:ZMSbar}
Z^{\overline {MS}} (\omega, \mu) = 1  + {107\over 105} \left(G_N m \omega\right)^2 \times \left[{1\over (d-4)_{UV}} +\gamma_E -\ln 4\pi\right],
\end{equation}
then Eq.~(\ref{eq:amp2eta2}) becomes
\begin{equation}
\label{eq:renormalized}
\left|{{\cal A}\over {\cal A}^R_{\eta^0}}\right|^2  = 1 + 2 \pi\left (G_N m \omega \right)  + \left (G_N m \omega \right)^2 \left[ -{214\over 105} \ln {{\bf k}^2\over 4\mu^2} + {4\pi^2\over 3} + {634913\over 44100}\right]  + {\cal O}(\eta^3),
\end{equation}
where now ${\cal A}^R_{\eta^0}$ is the leading order quadrupole amplitude, written in terms of the renormalized quadrupole $I^R_{ij}(\omega)$.   Note that, being a physical quantity, $|{\cal A}|^2$ should be independent of the arbitrary renormalization scale $\mu$.    Indeed from the the $\mu$ independence of the bare quadrupole $I^B_{ij}(\omega)$, together with Eq.~(\ref{eq:GRG}), Eq.~(\ref{eq:ZMSbar}), one finds a renormalization group (RG) equation
\begin{equation}
\label{eq:IRG}
\mu {d\over d\mu} I^R_{ij}(\omega,\mu) = - {214\over 105} (G_N m\omega)^2 I^R_{ij}(\omega,\mu),
\end{equation}
and thus from Eq.~(\ref{eq:amp2eta2}), the amplitude is $\mu$ independent, at least to order $\eta^2$.    Alternatively, Eq.~(\ref{eq:IRG}) could have also been derived less formally, by forcing the result of Eq.~(\ref{eq:amp2eta2}) to be scale independent, and introducing $\mu$ dependence of the renormalized moment in order to compensate for the explicit scale dependence of the logarithm.   

The RG equation for the quadrupole moment has the solution (dropping the superscript $R$ from now on)
\begin{equation}
\label{eq:RGsoln}
I_{ij}(\omega,\mu) = \left[{\mu\over \mu_0}\right]^{- {214\over 105} (G_N m\omega)^2} I_{ij}(\omega,\mu_0).
\end{equation}
Because of RG invariance of physical observables, any value of $\mu$ can be used in calculations.   In practice, it is useful to choose $\mu$ to minimize the logarithms in $|{\cal A}|^2$, so one can take $\mu\sim\omega$, where $\omega$ is the typical graviton frequency in the problem.    The scale $\mu_0$ should be taken to be of order the typical short distance scale, e.g. $\mu_0\sim a^{-1}$, where $a$ is the size of the system (the parameter that controls the multipole expansion).   At the level of the multipole EFT, $\mu_0$ is a free parameter, to be fit to data, and dependent on the details of the underlying gravitating system.    However, if the physics is known at the scale $\mu_0\sim a^{-1}$ where the multipole EFT begins to break down, it is possible to do an explicit \emph{matching} calculation to fix the precise value of $\mu_0$.   This is a standard procedure in effective field theories, see for example the reviews~\cite{IraEFT}.   We will come back to this issue in a separate paper.

In typical applications, the renormalization group is used to sum up the series of  large ``leading logarithms"  that arise whenever $\log \mu_0/\mu$ becomes large enough to compensate for the smallness of the expansion parameter, thereby improving naive perturbation theory.   Unfortunately, in gravitational wave physics the logarithms cannot become large.   This is because in the logarithmic terms $\eta^2 \ln \mu_0/\mu \sim \eta^2 \ln a/\lambda$ with $\eta \sim r_s / \lambda$, the extent of the gravitational wave source is bounded to be $a \ge r_s$ and therefore the logarithm can never compensate for the smallness of $\eta$.
For example, in the PN regime, one expects $\mu_0/\mu\sim v$, while the post-Minkowski expansion parameter becomes $\eta\sim v^3$, so that $\eta^2 \ln \mu_0/\mu \sim v^6\ln v$ is still small.   In this case, terms higher order than $(G_N m\omega)^2 \ln  \mu_0/\mu$ in Eq.~(\ref{eq:RGsoln}) are no larger than uncomputed PN corrections, and need not be resummed.
Nevertheless, the resummation of UV logarithms from the RG is systematic and can be used to obtain resummed gravitational wave observables in a factorized form, including instantaneous observables such as the waveform.

While summing the full series of logs in Eq.~(\ref{eq:RGsoln}) is typically not necessary for phenomenology, the RG equation for the quadrupole does contain information about the dynamics, as it constrains the pattern of logarithms can appear in the amplitude squared at higher orders in the calculation of quadrupole radiation from compact systems,
\begin{eqnarray}
\nonumber
\left|
{
{\cal A}(\omega)\over {\cal A}_{\eta^0}(\omega,\mu_0)
}
\right|_{\text{leading log}}^2 &=& 1 - {428\over 105} (G_N m \omega)^2 \ln {\omega\over \mu_0} + {91592\over 11025} (G_N m \omega)^4  \ln^2 {\omega\over \mu_0}\\
& & {} - {39201376\over 3472875}  (G_N m \omega)^6\ln^3{\omega\over \mu_0}+\cdots. \label{eq:UVlogsseries}
\end{eqnarray}
This set of terms is \emph{universal}, independent of the short distance dynamics.    It is a prediction of the RG that this series of terms will appear in long wavelength gravitational radiation from \emph{any} system.  To our knowledge, this series of terms, with the precise coefficients given above, has not been given previously in the literature.    It should be possible to test this prediction in the extreme mass limit, by expanding out the solutions of the Regge-Wheeler equation to the required order.   While Eq.~(\ref{eq:RGsoln}) does not seem to have immediate phenomenological applications, it may be possible to use this equation, for instance,  to construct phenomenological templates for gravitational wave emission, along the lines of refs.~\cite{approximants}.

We have referred to the series of logarithms in Eq.~(\ref{eq:UVlogsseries}) as the ``leading logarithms".  This means that this series captures the set of UV divergences of the form $\eta^{2n}\ln^n\Lambda$, $n\geq 1$, where $\Lambda\rightarrow\infty$ is a UV cutoff on integration momenta.      By power counting arguments similar to those applied in the previous section, one can show that there are new UV divergences, proportional to  $\ln\Lambda$, at every even order in the $\eta$ expansion.    These divergences can again be rendered finite by a renormalization of the multipole moments, with the consequence that the RG equation Eq.~(\ref{eq:IRG}) has the following structure to all orders in $\eta$,
\begin{equation}
\mu {d\over d\mu} I^R_{ij}(\omega,\mu) =\left[\sum^\infty_{n=1} \beta_{2n} (G_N m\omega)^{2n}\right] I^R_{ij}(\omega,\mu) ,
\end{equation}
where $\beta_{2}=-214/105$, and $\beta_{2n\geq 4}$ is obtained by computing UV poles in diagrams with $2n$ mass insertions.    Formally keeping only the contribution due to $\beta_2$ then sums the leading logs, keeping $\beta_4$ fixes the coefficients of all next-to-leading logs of the form $\eta^{2n}\ln^{n-1}\Lambda$ for all $n\geq 2$, and so on.   Note that this discussion neglects UV divergences in Feynman graphs with insertions of spin couplings.   These may introduce additional UV poles at odd orders in $\eta$.

\subsection{IR structure at higher orders}
\label{sec:IRsum}
\begin{figure}[!t]
\centerline{{\includegraphics{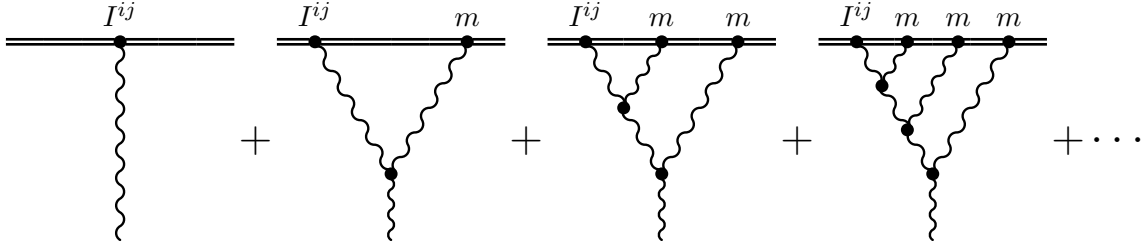}}}
\caption[1]{The series of leading IR poles.} \label{fig:ladders}
\vskip 0.5cm
\end{figure}

We have only shown the explicit cancellation of IR divergence up to order $\eta^2$, but clearly it should persist at higher orders.   A relatively simple class of such divergences is presented in Fig.~\ref{fig:ladders}.   These graphs contain the most singular IR divergence at each order in $\eta$.      We have explicitly computed the first two of these graphs in the previous section and shown them to have $1/(d-4)_{IR}$ and $1/(d-4)^2_{IR}$ divergences.   It is not difficult to show that this pattern persists at higher orders.

Consider a graph with similar topology to those in Fig.~\ref{fig:ladders}, with $n$ insertions of the mass monopole.    Such a graph contains integrations over $n$ spatial momenta ${\bf q}_{i=1,\ldots,n}$.   By examining the region of momentum  where all ${\bf q}_i\rightarrow 0$, one finds the behavior
\begin{equation}
{i\cal A}_n \rightarrow  i{\cal A}_{\eta^0} \times {1\over n!} \left[-32\pi (G_N m {\bf k}^2)\int {d^{d-1}{\bf q}\over (2\pi)^{d-1}} {1\over {\bf q}^2} {1\over 2 {\bf k}\cdot {\bf q}}\right]^n.
\end{equation}
Then, using the result 
\begin{equation}
\int {d^{d-1}{\bf q}\over (2\pi)^{d-1}} {1\over {\bf q}^2} {1\over 2 {\bf k}\cdot {\bf q}-i\epsilon} = -{i\over 16\pi |{\bf k}|} {1\over (d-4)_{IR}}  ,
\end{equation}
one finds that the leading IR poles sum into
\begin{equation}
\sum_{n=0}^\infty  {\cal A}_n \rightarrow {\cal A}_{\eta^0} \exp\left[{2 i G_N m \omega \over (d-4)_{IR}}\right].
\end{equation}
The IR divergences have summed into a harmless overall phase factor, and therefore cancel from physical quantities.   The summation of IR divergences due to Coulomb-type phases in soft graviton emission processes was first discussed by Weinberg~\cite{Weinberg:1965nx}.   

Associated with the leading IR divergences are additional finite parts, which do contribute to $|{\cal A}|^2$.     These contributions come with factors of $\pi$ at every order in $\eta$, and tend to be numerically enhanced.   The precise form of these corrections can be obtained by solving the wave equation for the propagation of a graviton in the presence of the $1/r$ gravitational field of the source, see refs.~\cite{Khriplovich:1997ms, Asada:1997zu}.     Their result is that the leading order amplitude squared  $|{\cal A}_{\eta^0}|^2$ gets scaled by the Sommerfeld factor   
\begin{equation}
S(\omega ) = \frac{4 \pi G_N m \omega}{1 - \exp\left(- 4 \pi G_N m  \omega \right)}  \label{eq:sommerff}.
\end{equation}
This result has also been employed in \cite{Damour:2007xr} in the construction of phenomenological waveforms.

Taken together, the resummation of the leading UV logarithms of Eq. (\ref{eq:UVlogsseries}) and the resummation of the leading IR effects via Eq. (\ref{eq:sommerff}) predict the existence of a term at order $\eta^3$ that has the form
\begin{equation}
\left|{{\cal A}\over{\cal A}_{\eta^0}}\right|_{\eta^3} \supset- {856 \pi \over 105} (G_N m \omega )^3 \ln {\omega \over \mu_0}.
\end{equation}
This prediction agrees with results for extreme mass ratio post-Newtonian systems, calculated by the methods reviewed in~\cite{BHpert}.

\section{The multipole expansion}
\label{sec:mult}

It remains to match onto the multipole EFT of Eq.~(\ref{eq:lag}), that is, to compute the specific form of the moments given the details of the short distance physics.   In order to match, one must compute graviton emission amplitudes in the full theory (containing details about short distance structure), expanded in the long wavelength limit, and compare them to Eq.~(\ref{eq:lag}).   One then adjusts the moments  in Eq.~(\ref{eq:lag}) in order to obtain agreement between the two calculations.

For classical processes, it is often sufficient to consider single graviton emission amplitudes.  These are generated in the full theory by the interaction
\begin{equation}
\label{eq:tadpole}
\Gamma[\bar h] = -{1\over 2 m_{Pl}}\int d^4 x T^{\mu\nu}(x) {\bar h}_{\mu\nu},
\end{equation}
where the tensor $T^{\mu\nu}(x)$ should be regarded as a Minkowski spacetime tensor, measured relative to a Lorentz frame asymptotically far from the source.   $T^{\mu\nu}(x)$ is defined in terms of the off-shell amplitude for single graviton emission
\begin{equation}
{\cal A}_{\mu\nu} = -{1\over 2 m_{Pl}} \int d^4 x T_{\mu\nu}(x) e^{i k\cdot x},
\end{equation}
and therefore satisfies the conservation law $\partial_\mu T^{\mu\nu}(x)=0$ on account of the Ward identity for graviton amplitudes.   The tensor  $ T^{\mu\nu}(x)$ can therefore be regarded as the energy-momentum ``pseudo-tensor'' for gravity plus matter that appears in all approaches to gravitational radiation.

The multipole expansion is generated by taking the limit ${\bf k}\rightarrow 0$, or in coordinate space, by expanding Eq.~(\ref{eq:tadpole}) as 
\begin{equation}
{\bar h}_{\mu\nu}(x) = \sum_{n=0}^\infty {1\over n!} {\bf x}^{i_1} \cdots {\bf x}^{i_n} \partial_{i_1} \cdots \partial_{i_n} {\bar h}_{\mu\nu}(x^0,0),
\end{equation}
where the point ${\bf x}=0$ is taken to be the center of mass, defined such that 
\begin{equation}
{\bf X}^i = \int d^3{\bf x} T^{00} {\bf x}^i=0.
\end{equation}
 This expansion can be truncated at finite order whenever the condition ${\bf x}\cdot \nabla\sim a/\lambda\ll 1$ holds.
Plugging into Eq.~(\ref{eq:tadpole}) and breaking up the moments
\begin{equation}
\int d^3{\bf x} T^{\mu\nu} {\bf x}^{i_1} \cdots {\bf x}^{i_n},
\end{equation}
into irreducible representations of the rotation group gives the result Eq.~(\ref{eq:lag}), at linear order in the radiation field ${\bar h}_{\mu\nu}$.    The non-linear terms,  usually not needed for classical applications, then follow by covariantizing the linear result.

The decomposition of moments of $T^{\mu\nu}$ into irreducible representations is standard in the general relativity literature, see for example~\cite{maggiore}. Here we recall some of the manipulations at low orders in the expansion.   To zeroth order in derivatives, the multipole expansion just gives
\begin{equation}
\label{eq:LOmult}
\Gamma[{\bar h}]_0= -{1\over 2m_{Pl}}\int dx^0 m {\bar h}_{00}(x^0,0),
\end{equation}
where $m=\int d^3 {\bf x} T^{00}$ is the total energy of the isolated system. The relation $\partial_\mu T^{\mu\nu}=0$ implies that this is a conserved quantity, ${\dot m}=0$.    Eq.~(\ref{eq:LOmult}), of course reproduces the mass monopole term $-m\int d\tau$ in Eq.~(\ref{eq:lag}),  to linear order in the fields.     We have not included in  Eq.~(\ref{eq:LOmult}) the contribution from
\begin{equation}
\label{eq:momentum}
\int dx^0\left[\int d^3 {\bf x} T^{0i}(x^0,{\bf x})\right] {\bar h}_{0i}(x^0,0),
\end{equation}
since by conservation of $T^{\mu\nu}$, 
\begin{equation}
{\bf P}^i = \int d^3 {\bf x} T^{0i}(x^0,{\bf x}) = \int d^3{\bf x} {\dot T}^{00}(x^0,{\bf x}) {\bf x}^i \equiv m {\dot{\bf X}}^i=0,
\end{equation}
in the center of mass frame.   Even if this term were non-zero, for on-shell radiation $\partial_0\sim \partial_i$, so this coupling to ${\bar h}_{0i}$ would not appear until first order in derivatives.   Likewise the coupling 
\begin{equation}
\int dx^0\left[\int d^3 {\bf x} T^{ij}(x^0,{\bf x})\right] {\bar h}_{ij}(x^0,0),
\end{equation}
does not appear until second order in derivatives, given the relation
\begin{equation}
\int d^3 {\bf x} T^{ij} (x^0,{\bf x})= {1\over 2} \, \partial_0^2 \int d^3{\bf x} T^{00} (x^0,{\bf x}){\bf x}^i {\bf x}^j.
\end{equation}
which follows from energy-momentum conservation.

Including all terms that survive at first order in derivatives, one finds
\begin{equation}
\label{eq:NLOmult}
\Gamma[{\bar h}]_1 =- \frac{1}{2} \int dx^0  L^{ij}\omega^{ij}_0.
\end{equation}
The linearized spin connection is
\begin{eqnarray}
\omega^{ij}_\mu &=&{1\over 2 m_{Pl}} \left(\partial^i {{\bar h}^j}_\mu - \partial^j {{\bar h}^i}_\mu\right),
\end{eqnarray}
and the angular momentum is given $L^{ij}= - \int d^3 {\bf x} \left(T^{0i} {\bf x}^j - T^{0j} {\bf x}^i\right)$.   This agrees with the terms $-\frac{1}{2}\int dx^\mu L_{ab} \omega^{ab}_\mu$  in Eq.~(\ref{eq:lag}), again to linear order in the fields.   Note that by $\partial_\mu T^{\mu\nu}=0$, the angular momentum is conserved ${\dot L}_{ij}=0$.

Multipoles which source radiation start appearing at second order in derivatives.  At this point, it is convenient to split up  the different multipoles into radiative terms and non-radiative terms.   The radiative terms couple to ${\bar h}_{ij}$ and therefore source physical radiation.   The non-radiative terms are couplings either to constants of the motion, as in Eq.~(\ref{eq:LOmult}), Eq.~(\ref{eq:NLOmult}), to quantities that can be made to vanish by a choice of inertial frame, or couplings to the Ricci tensor and its covariant derivatives.  On account of the vacuum field equations, $R_{\mu\nu}=0$,  the latter can be made to vanish by field redefinitions.   

From now on we will omit all non-radiative terms.   To work out the radiative couplings, it is sufficient to set ${\bar h}_{00}={\bar h}_{0i}=0$ and retain only the physical radiation field ${\bar h}_{ij}$.    As discussed above, there is one term at second order in derivatives of the form ($TF$ stands for the traceless part of the tensor)
\begin{equation}
\Gamma[{\bar h}]_2 = {1\over 2} \int dx^0 \left[\int d^3 {\bf x}  T^{00} {\bf x}^i {\bf x}^j\right]^{TF} E_{ij},
\end{equation}
with the linearized electric parity Weyl tensor for on-shell 
\footnote{Had we kept terms with ${\bar h}_{00}$ and ${\bar h}_{0i}$ we would have found the same result, but with the full expression $E_{ij}\rightarrow (\partial_0 \partial_i{\bar h}_{0j}+ \partial_0 \partial_j{\bar h}_{0i}- \partial_i \partial_j {\bar h}_{00}  - \partial_0^2 {\bar h}_{ij})/(2 m_{Pl})$, as expected from gauge invariance of the action.} ${\bar h}_{ij}$,
\begin{equation}
E_{ij} = -{1\over 2 m_{Pl}} \partial^2_0{\bar h}_{ij}.
\end{equation}

There are two additional contributions at second order in spacetime derivatives.   One arises from the expansion of ${\bar h}_{ij}$ to first order in spatial gradients, and is proportional to the moment $\int d^3 {\bf x} T^{ij} {\bf x}^k,$ which decomposes into $SO(3)$ irreducible representations as
\begin{equation}
\label{eq:CG1}
({\bf 2}\oplus {\bf 0})\otimes {\bf 1} \sim {\bf 3}_E\oplus {\bf 2}_M\oplus{\bf 1}\oplus {\bf 1}.
\end{equation}
The representations ${\bf 2}$, ${\bf 3}$ correspond to magnetic quadrupole and electric octupole moments, while the two ${\bf 1}$ representations are non-radiative, and vanish by the use of the equations of motion.   Of the radiative terms, the ${\bf 2}$ contributes at two-derivative order.   Carrying out standard manipulations which will not be repeated here (see, e.g., the textbook~\cite{maggiore}) one finds a contribution
\begin{equation}
\left[\int d^3{\bf x} T^{ij} {\bf x}^k \partial_k {\bar h}_{ij} \right]_{{\bf 2}_M}  = - \frac{4}{3} J^{ij} B_{ij},
\end{equation}
where we have defined the leading order magnetic quadrupole,
\begin{equation}
\label{eq:magquad}
J^{ij}   = - \frac{1}{2} \int d^3 {\bf x} \left(\epsilon^{ikl} \left[T^{0k} {\bf x}^j {\bf x}^l\right]^{TF} + \epsilon^{jkl} \left[T^{0k} {\bf x}^i {\bf x}^l\right]^{TF}\right).
\end{equation}
The magnetic parity Weyl tensor is
\begin{equation}
B_{ij} = \frac{1}{2} \epsilon_{imn} C_{0jmn}  = \frac{1}{2 m_{Pl}} \epsilon_{imn} \left(\partial_0 \partial_n \bar h_{jm} + \partial_j \partial_m \bar h_{0n} \right).
\end{equation}
The remaining contribution at two derivatives is generated by the term in the expansion of ${\bar h}_{ij}$ containing two spatial gradients.  This couples to $\int d^3 {\bf x} T^{ij} {\bf x}^k {\bf x}^l,$ whose decomposition is
\begin{equation}
\label{eq:CG2}
({\bf 2}\oplus {\bf 0})\otimes ({\bf 1}\otimes {\bf 1})_S\sim {\bf 4}_E \oplus {\bf 3}_M \oplus {\bf 2}_E\oplus{\bf 2}\oplus {\bf 2}\oplus {\bf 1}\oplus {\bf 0}\oplus {\bf 0}.
\end{equation}
It turns out that the last five moments couple to the Ricci curvature, and are therefore irrelevant for radiation.   However, the term
\begin{eqnarray}
\label{eq:2E}
\nonumber
\left[{1\over 2}\int d^3 {\bf x} T^{ij} {\bf x}^k {\bf x}^l \partial_k\partial_l {\bar h}_{ij}(x^0,0)\right]_{{\bf 2}_E} &=  \int  d^3{\bf x} \left[\frac{11}{42} T^{ij} {\bf x}^2 + \frac{2}{21} T^{kk} {\bf x}^i {\bf x}^j \right.\\
&  \left.- \frac{1}{7} T^{ik} {\bf x}^j {\bf x}^k - \frac{1}{7} T^{jk} {\bf x}^i {\bf x}^k \right]^{TF}\partial_0^2 {\bar h}_{ij}.
\end{eqnarray}
has parts which contribute to the two-derivative action.   One can simplify Eq.~(\ref{eq:2E})  by applying $\partial_\mu T^{\mu\nu}=0$, together with integration by parts, with the result
\begin{eqnarray}
\nonumber
\left[{1\over 2}\int d^3 {\bf x} T^{ij} {\bf x}^k {\bf x}^l \partial_k\partial_l {\bar h}_{ij}(x^0,0)\right]_{{\bf 2}_E} =  {1\over 2} \int  d^3{\bf x} \left(T^{kk} -{4\over 3} {\dot T}^{0k} {\bf x}^k + {11\over 42} {\ddot T}^{00} {\bf x}^2\right) \left[{\bf x}^i{\bf x}^j\right]^{TF} \partial_0^2 {\bar h}_{ij}.\\
\end{eqnarray}
This contains the second derivative term we are after, together with higher order three- and four- derivative terms.    Putting everything together, we find at second order in derivatives
\begin{equation}
\Gamma[{\bar h}]_2 =   \frac{1}{2} \int dx^0 \left(I^{ij} E_{ij} - \frac{4}{3} J^{ij} B_{ij}\right),
\end{equation}
with 
\begin{equation}
I^{ij} = \int d^3 {\bf x} (T^{00} + T^{kk}) \left[{\bf x}^i{\bf x}^j\right]^{TF} + {\cal} O(\partial^1),
\end{equation}
and $J^{ij}$ given in Eq.~(\ref{eq:magquad}).

The pattern is similar at higher orders.  At third order in the derivative expansion, there are contributions from the ${\bf 3}_E$ in Eq.~(\ref{eq:CG1}), from ${\bf 3}_M$ in Eq.~(\ref{eq:CG2}), and from the coupling to $\int d^3{\bf x} T^{ij} {\bf x}^k {\bf x}^l {\bf x}^m$, which induces order $\partial^1$ corrections to the magnetic quadrupole and one of the leading parts of the electric octupole moment.    The terms in the quadrupole moment at order $\partial^n$ arise from the expansion of  ${\bar h}_{ij}$ in spatial gradients at orders $n+2$, $n+1$ and $n$.  More generally, the set of terms in an electric parity $\ell$-pole moment at order $\partial^n$ receive contributions from the expansion of  ${\bar h}_{ij}$ in spatial gradients at orders $n+\ell$, $n+\ell - 1$ and $n + \ell - 2$. Magnetic parity $\ell$-poles at order $\partial^n$ receive contributions from the expansion of  ${\bar h}_{ij}$ in spatial gradients at orders $n+\ell+1$, $n+\ell$ and $n + \ell - 1$.

Summarizing our results at low orders in the expansion, we find for the first three moments which can radiate,
\begin{equation}
\Gamma[{\bar h}]= {1\over 2} \int dx^0 \left(I^{ij} E_{ij} - \frac{4}{3} J^{ij} B_{ij} + \frac{1}{3} I^{ijk} \partial_k E_{ij} +\cdots \right),
\end{equation}
with
\begin{eqnarray}
\label{eq:I2d}
I^{ij} &=& \int d^3 {\bf x} \left(T^{00} + T^{kk}-{4\over 3} {\dot T}^{0k} {\bf x}^k + {11\over 42} {\ddot T}^{00} {\bf x}^2\right) \left[{\bf x}^i{\bf x}^j\right]^{TF}  + {\cal O}(\partial^2),\\
I^{ijk} &=& \int d^3 {\bf x} (T^{00}+ T^{ll}) \left[{\bf x}^i {\bf x}^j {\bf x}^k\right]^{TF} + {\cal O}(\partial^1),\\
J^{ij} &=& - \frac{1}{2} \int d^3 {\bf x} \left(\epsilon^{ikl} T^{0k} {\bf x}^j {\bf x}^l + \epsilon^{jkl} T^{0k} {\bf x}^i {\bf x}^l \right) + {\cal O}(\partial^1).
\end{eqnarray}
Given a model for the gravitating source (i.e., a prescription for computing the off-shell emission amplitude $T^{\mu\nu}$), these formulas and their generalization to higher orders in derivatives can be used to compute long wavelength radiation observables.  The results presented here are sufficient for computing post-Newtonian radiation to order $v^2$ beyond the leading order quadrupole formula, as we discuss in the next section.

\section{Applications to post-Newtonian systems}
\label{sec:PN}

In order to apply the general formalism developed above to post-Newtonian systems, one must compute the amplitude $T^{\mu\nu}(x)$ in terms of the orbital degrees of freedom of the non-relativistic compact system.  We will only consider spinless systems in this paper and leave the issue of spin for future work~\cite{PRR}.     

In momentum space,  $T^{\mu\nu}(k)$ can be calculated as a sum over Feynman diagrams with a single external (off-shell) graviton of momentum $k^\mu$.    By definition, external propagators are stripped off.   For PN systems it is convenient to perform this calculation using NRGR~\cite{GnR1}, which is tailored for non-relativistic gravitating bound states.   In this EFT, the gravitational degrees of freedom are decomposed into potential modes $H_{\mu \nu}$ with support over the size $r$ of the bound state, and on-shell radiation modes ${\bar h}_{\mu\nu}$.   The NRGR radiation mode exactly coincides with the graviton field appearing in the multipole EFT written down in Eq.~(\ref{eq:lag}).   The potential mode, being a short distance field, must be integrated out.    In diagrammatic language, this means simply that $T^{\mu\nu}(k)$ is computed by considering only graphs that contain internal potential graviton propagators, with a single external radiation line of momentum $k^\mu$.    The relevant Feynman rules are given in~\cite{GnR1,GnR1.5}.  In particular the propagator for the potential graviton is
\begin{equation}
\langle {H_{\bf q}}_{\mu\nu}(x^0)  {H_{\bf k}}_{\mu\nu}(0)\rangle = -{i\over {\bf q}^2} (2\pi)^3\delta({\bf q}+{\bf k}) \delta(x^0) P_{\mu\nu;\alpha\beta},
\end{equation}
where in the gauge employed in~\cite{GnR1}, $P_{\mu\nu;\alpha\beta}={1\over 2}\left[\eta_{\mu\alpha} \eta_{\nu\beta} +\eta_{\mu\beta} \eta_{\nu\alpha} -{2\over d-2} \eta_{\mu\nu} \eta_{\alpha\beta}\right]$.

To illustrate how this works in practice, we now compute $T^{\mu\nu}(k)$ to next-to-leading order in the velocity expansion.   We stress, however, that the procedure is well defined to all orders in the expansion parameter $v$.   To be definite, we consider an ensemble of gravitating point particles with masses $m_a$ and positions ${\bf x}_a(t)$, with $a$ labeling the particle species.  Introducing the partial Fourier transform $T^{\mu\nu}(x^0,{\bf k})=\int d^3 {\bf x} e^{-i{\bf k}\cdot {\bf x}} T^{\mu\nu}(x^0,{\bf x}),$ we can read off the moments from the Taylor expansion about ${\bf k}=0,$
\begin{equation}
T^{\mu\nu}(x^0,{\bf k}) = \sum_{n=0}^\infty {(-i)^n\over n!} \left(\int d^3 {\bf x} T^{\mu\nu}(x^0,{\bf x}) \, {\bf x}^{i_1}\cdots {\bf x}^{i_n}\right) {\bf k}_{i_1} \cdots {\bf k}_{i_n}.
\end{equation}
By construction, in NRGR, each term in this expansion corresponds to a sum of Feynman graphs that scale as a definite power of the expansion parameter $v\ll 1$.

\begin{figure}[!t]
\centerline{\scalebox{0.99}{\includegraphics{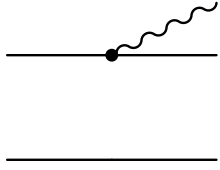}}}
\caption[1]{Feynman diagram yielding the leading terms for $T^{00}$ and $T^{0i}$.} \label{fig:T000iLO}
\vskip 0.5cm
\end{figure}

To leading order in velocity, the components of $T^{\mu\nu}(x^0,{\bf k})$ can be obtained from the diagram in Fig. \ref{fig:T000iLO}.   The results are
\begin{equation}
T^{00}(x^0,{\bf k})= \sum_a m_a e^{-i {\bf k}\cdot {\bf x}_a},
\end{equation}
which generates the moments  
\begin{eqnarray}
m       &=&  \sum_a m_a + {\cal O}(v^2)\\
{\bf X}^i  &=&  \frac{1}{M} \sum m_a {\bf x}^i_a + {\cal O}(v^2)\\
I^{ij}   &=&  \sum_a  m_a \left[{\bf x}^i_a {\bf x}^j_a\right]^{TF} + {\cal O}(v^2)\\
I^{ijk}   &= & \sum_a m_a \left[{\bf x}_a^i {\bf x}_a^j {\bf x}_a^k\right]^{TF} + {\cal O}(v^2) \label{eq:octo} \\ 
\int d^3 {\bf x} T^{00} {\bf x}^2 \left[{\bf x}^i {\bf x}^j\right]^{TF} &=&  \sum_a m_a {\bf x}_a^2 \left[{\bf x}^i_a {\bf x}^j_a\right]^{TF}
\end{eqnarray}
where $M = \sum_a m_a$. At order $v$ beyond the leading order we get the first contribution to $T^{0i}$,
\begin{equation}
T^{0i}(x^0,{\bf k})  = \sum_a m_a {\bf v}_a^i e^{-i {\bf k}\cdot {\bf x}_a},
\end{equation}
which to this order in $v$ yields 
\begin{align}
{\bf P}^i    & =\sum_a m_a {\bf v}_a^i + \mathcal O(v^3) \\
{\bf L}_i    & =\epsilon_{ijk} L^{jk} =\sum_a m_a ({\bf x}_a \times {\bf v}_a)^i + \mathcal O(v^3)\\
J^{ij} & = - \frac{1}{2} \sum_a m_a \left(({\bf v}_a\times {\bf x}_a)^i {\bf x}^j_a + ({\bf v}_a\times {\bf x}_a)^j {\bf x}^i_a\right) + \mathcal O(v^3) \label{eq:cquad} \\
\int d^3{\bf x} T^{0k} {\bf x}^k \left[{\bf x}^i {\bf x}^j\right]^{TF} & = \sum_a m_a {\bf v}_a \cdot {\bf x}_a \left[{\bf x}^i_a {\bf x}^j_a\right]^{TF}
\end{align}

\begin{figure}[!t]
\centerline{\scalebox{0.99}{\includegraphics{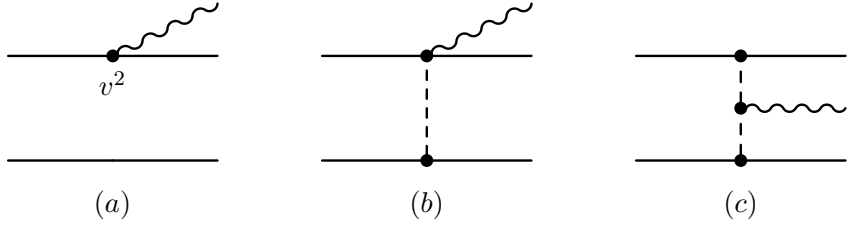}}}
\caption[1]{Diagrams for $v^2$ corrections to $T^{00}$. Dashed lines denote potential gravitons.} \label{fig:T00NLO}
\vskip 0.5cm
\end{figure}

\begin{figure}[!t]
\centerline{\scalebox{0.99}{\includegraphics{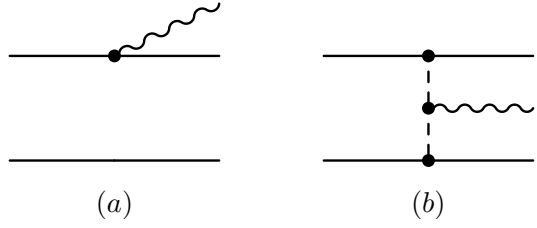}}}
\caption[1]{Diagrams needed to compute the leading contribution to $T^{ij}$.} \label{fig:TijLO}
\vskip 0.5cm
\end{figure}

At order $v^2$ beyond the leading order, there are corrections to $T^{00}$ from the graphs in Fig.~\ref{fig:T00NLO} as well as new contributions to $T^{ij}$ from the diagrams in Fig. \ref{fig:TijLO}.  For $T^{00}$ the results are
\begin{eqnarray}
\mbox{Fig. \ref{fig:T00NLO}(a)} &=& {1\over 2} \sum_a m_a {\bf v}^2_ae^{-i {\bf k}\cdot {\bf x}_a}\\
\mbox{Fig. \ref{fig:T00NLO}(b)} &=& {i\over 4 m_{Pl}^2} \sum_{a,b} m_a m_b \int_{\bf q} e^{i{\bf q}\cdot {\bf x}_{ab}}{- i\over 2 {\bf q}^2} e^{-i {\bf k}\cdot {\bf x}_a} = \sum_{a,b} {G_N m_a m_b\over|{\bf x}_{a}-{\bf x}_b|} e^{-i {\bf k}\cdot {\bf x}_a}\\
\mbox{Fig. \ref{fig:T00NLO}(c)}   &=& {1\over 2!} \frac{- i}{2 m_{Pl}} \sum_{a,b}{m_a m_b} \int_{\bf q} e^{i{\bf q}\cdot {\bf x}_{ab} } {- i\over 2 {\bf q}^2} {- i\over 2 ({\bf q} + {\bf k})^2} {3 i\over m_{Pl}} \left(\mathbf q^2 + \mathbf q \cdot \mathbf k\right) e^{-i{\bf k}\cdot {\bf x}_b} \nonumber \\
&=& - \frac{3}{2} \sum_{a,b} {G_N m_a m_b\over|{\bf x}_a-{\bf x}_b|} e^{-i {\bf k}\cdot {\bf x}_a}
\end{eqnarray}
with $\int_{\bf q} = \int \frac{d^3\bf q}{(2 \pi)^3}$ and where we have used the ${\bar h} H H$ Feynman rule derived in ref.~\cite{GnR1}.   To order $v^2$ we need the contributions of these graphs up to second order in the multipole expansion.   This gives 
\begin{eqnarray}
m &=& \sum_a {\bar m}_a +{\cal O}(v^4),\\
\label{eq:CMX}
{\bf X}^i &=& \frac{1}{M} \sum_a {\bar m}_a {\bf x}^i_a+{\cal O}(v^4),\\
\int d^3{\bf x} T^{00} \left[{\bf x}^i {\bf x}^j\right]^{TF}  &=& \sum_a {\bar m}_a \left[{\bf x}^i_a {\bf x}^j_a \right]^{TF} +{\cal O}(v^4),
\end{eqnarray}
where we have defined a ``renormalized'' mass
\begin{equation}
{\bar m}_a = m_a\left[1 + {1\over 2} {\bf v}^2_a -{1\over 2} \sum_b {G_N m_b\over |{\bf x}_a -{\bf x}_b|}\right].\end{equation}

To complete the full set of order $v^2$ corrections to the mass quadrupole $I^{ij}$, we also need to compute the leading order moments of $T^{kk}$. We find
\begin{eqnarray}
\mbox{Fig. \ref{fig:TijLO}(a)} &=&  \sum_a m_a {\bf v}^2_ae^{-i {\bf k}\cdot {\bf x}_a},\\
\mbox{Fig. \ref{fig:TijLO}(b)} &=& {1\over 3}\times \mbox{Fig. \ref{fig:T00NLO}(c)},
\end{eqnarray}
and therefore 
\begin{equation}
\int d^3{\bf x} T^{kk} \left[x^i x^j\right]^{TF} = \sum_a m_a \left({\bf v}^2_a-{1\over 2} \sum_b {G_N m_b\over |{\bf x}_a -{\bf x}_b|}\right) \left[{\bf x}^i _a {\bf x}^j_a\right]^{TF}.
\end{equation}
Combining all of our results we finally obtain, from Eq.~(\ref{eq:I2d}),
\begin{eqnarray}
\nonumber
I^{ij} &=& \sum_a m_a \left(1 + {3\over 2} {\bf v}^2_a -  \sum_b {G_N m_b\over |{\bf x}_a -{\bf x}_b|}\right) \left[{\bf x}^i_a {\bf x}^j_a\right]^{TF} + {11\over 42}\sum_a m_a {d^2\over dt^2} \left({\bf x}_a^2 \left[{\bf x}^i_a {\bf x}^j_a\right]^{TF}\right)\\
& & {}-{4\over 3}\sum_a m_a {d\over dt} \left({\bf x_a}\cdot {\bf v}_a \left[{\bf x}^i_a {\bf x}^j_a\right]^{TF}\right) + {\cal O}(v^4) \label{eq:equadv2}.
\end{eqnarray}

It is straightforward to use the results obtained in this section to compute corrections to the quadrupole radiation formula.    For illustration we consider two particles in a circular orbit of frequency $\Omega$, and work in the center of mass frame, ${\bf X}=0$, with ${\bf X}$ given in Eq.~(\ref{eq:CMX}).    Using the 1PN equations of motion, we obtain
\begin{align}
I^{ij}(t)& =  \mu \left\{1 -\left(\frac{1}{42} + \frac{39}{42} \nu \right) x \right\} \left[{\bf x}^i {\bf x}^j \right]^{TF} + \frac{11}{21} \mu {\bf x}^2 \left(1 - 3 \nu\right) \left[{\bf v}^i {\bf v}^j \right]^{TF}  \label{eq:QuadCirc},\\
J^{ij}(t) & = - \frac{1}{2} \mu \sqrt{1 - 4 \nu} \left[({\bf v} \times {\bf x})^i {\bf x}^j + ({\bf v} \times {\bf x})^j {\bf x}^i\right] \label{eq:CQuadCirc},\\
I^{ijk}(t) & = \mu \sqrt{1 - 4 \nu} \left[{\bf x}^i {\bf x}^j {\bf x}^k \right]^{TF} \label{eq:OctoCirc},
\end{align}
with ${\bf x}={\bf x}_1-{\bf x}_2$ the relative coordinate, ${\bf v}=\dot{\bf x}$, and  $\mu={m_1 m_2}/{M}$, $\nu = \mu/M$.     The PN expansion parameter is $x = (G_N M \Omega)^{2/3}$. 
The square modulus of the frequency domain moments (dropping terms involving $\delta(\omega)$ which do not contribute to radiation),
\begin{eqnarray}
\left| I^{ij}(\omega)\right|^2 &=& {\pi T \mu^2  x^2\over 2 \Omega^4}  \left[1+ \left(-{107\over 21}+  {55\over 21}\nu\right)x\right]  \left[ \delta(\omega-2\Omega) + \delta(\omega+2\Omega)\right],\\
\left| J^{ij}(\omega)\right|^2 &=& {\pi T \mu^2  x^3\over 2 \Omega^4}(1-4\nu) \left[ \delta(\omega-\Omega) + \delta(\omega+\Omega)\right] ,\\
\nonumber
\left| I^{ijk}(\omega)\right|^2 &=& {\pi T \mu^2  x^3\over 4 \Omega^6} (1-4\nu) \left[ \delta(\omega-3\Omega) + \delta(\omega+3\Omega)+  {3\over 5}\left(\delta(\omega-\Omega) + \delta(\omega+\Omega)\right)\right],\\
\end{eqnarray}
with $T=2\pi\delta(0)$, can then be plugged into Eq.~(\ref{eq:PowerIJIcoo}), yielding the well known 1PN result
\begin{equation}
{{\dot P}^0 \over  {\dot P}^0_{LO}} =1 - \left(\frac{1247}{336} + \frac{35}{12} \nu \right) x.
\end{equation}

Combining the results of this section with those of sec.~\ref{sec:postM}, we can also obtain the 1.5PN and 2.5PN radiative (or ``tail") corrections to the emitted power.   This follows from Eq.~(\ref{eq:PowerIJImom}) and Eq.~(\ref{eq:eta1tail}) together with the 1PN mass monopole 
\begin{equation}
{m\over M} =1 - \frac{\nu}{2} x.
\end{equation}
One obtains,
\begin{equation}
{{\dot P}_{tail}^0 \over  {\dot P}^0_{LO}} = 4 \pi x^{3/2} - \left(\frac{8191}{672} + \frac{583}{24} \nu\right)  \pi x^{5/2}+\cdots.
\end{equation}
Here we have used the universality of the order $\eta^1$ tail correction to obtain the coefficient of $x^{5/2}$.  Finally, including the results of sec.~\ref{sec:eta2}, we also obtain the term at 3PN that is non-analytic in the expansion parameter $x$
\begin{equation}
\left.{{\dot P}_{na}^0 \over  {\dot P}^0_{LO}}\right|_{x^3} = -{856\over 105} x^3\ln x,
\end{equation}
where the renormalization scale has been chosen as $\mu_0\sim 1/r$.

\section{Conclusions}
\label{sec:conclusions}

In this paper, we have extended the EFT approach to compact gravitating systems, first proposed in~\cite{GnR1}, to account for radiative corrections to gravitational wave emission.   This includes corrections that arise from the multipole expansion $a/\lambda\ll1$, and from corrections to vacuum wave propagation, suppressed by powers of $\eta=G_N m/\lambda$, that are present due to the non-linear nature of the Einstein equations.    Many of the results obtained in this paper, in particular those of sec.~\ref{sec:postM}, are universal, independent of the precise nature of the gravitating source.    They can be used not only for post-Newtonian systems (as we consider in sec.~\ref{sec:PN}) but to other physical situations involving low frequency gravitons, such as soft bremsstrahlung in relativistic black hole collisions or graviton absorption by black holes.

Of course, the EFT approach necessarily shares some features in common with existing approaches~\cite{postM1,postM2} to gravitational radiation from localized sources.    For example, the separation of scales made explicit by an EFT is physically equivalent to the standard matched asymptotic expansions introduced in~\cite{burke,postM1}.   However, there are some aspects of the problem of gravitational radiation that are particularly well suited to EFT methods, namely the issue of how to handle the UV divergences that appear at sufficiently high order in perturbation theory (order $\eta^2$ and beyond).   Indeed, we saw that in the EFT, it is relatively straightforward to identify and isolate the terms in the perturbative expansion that contain UV poles, and to renormalize these terms by identifying a suitable counterterm in the effective theory Lagrangian.  As we saw, this necessarily implies non-trivial RG flow of the multipole moments, which can be used to put constraints on the form of the non-analytic dependence on frequency of observables.    Likewise, the identification, resummation, and cancellation of IR divergences is made transparent in the way that the EFT computes perturbative corrections, as we discussed in sec.~\ref{sec:postM}.     What emerges from our analysis is the following pattern of non-analytic terms in observables associated with radiation of frequency $\omega$ from a fixed multipole
\begin{equation}
|{\cal A}_\ell(\omega)|^2 \sim S(\omega) \sum_{n=1}^\infty  \eta^{2 n}  L \left[\beta^{(\ell)}_{2n}  +  {\cal O}(\eta^{2} L)+ {\cal O}(\eta^{2} L)^2  + {\cal O}(\eta^{2} L)^3 +\cdots\right],
\end{equation}
where $S(\omega)$ is the Sommerfeld factor in Eq.~(\ref{eq:sommerff}), and $L=\ln a\omega$.   The coefficient $\beta_{2n}^{(\ell)}$ fixes the entire series of logs in the square brackets.

There are several issues that have we not dealt with, but which we expect should be tractable by the methods presented here.   For example, we have not included the effects of total angular momentum, which contribute beyond 3PN and were computed, using different methods, in~\cite{Blanchet:1997ji}.   It is clear that such terms are generated in our formalism by inclusion of the angular momentum coupling that appears in Eq.~(\ref{eq:lag}).   We also have not considered in this paper the ``memory" effect~\cite{memory,Wiseman:1991ss,Blanchet:1997ji} which arises from the self interactions of the gravitational radiation.    It corresponds to graphs with more than one coupling to multipole moments which can source radiation (quadrupole and higher) together with one or more graviton self-interactions.    We do not see any obstruction to computing the resulting Feynman integrals by methods similar to those employed here.   The matching to PN systems in sec.~\ref{sec:PN} has not been carried out to sufficiently high order to fix the matching scale $\mu_0$ in the RG equation for the quadrupole moment.   In principle, this computation is well defined in our approach, provided one regulates the theory consistently by dimensional regularization, as we have shown how to do here for the long distance $\eta$ corrections.   Finally, we have restricted ourselves to time averaged observables generated by $|{\cal A}|^2$.    Thus we did not consider issues such as the computation of instantaneous waveforms or the back reaction of radiation on the multipole moments that serve as inputs into Eq.~(\ref{eq:lag}).   It seems reasonable to imagine that some version of the formalism introduced in~\cite{galley1,galley2} could be applied to this problem.   We hope to examine some of these questions in future work.

\acknowledgments
We thank Ira Rothstein for his input, and for collaboration in the early stages of this project.  A. R. acknowledges the hospitality of Barak Kol and his group at the Racah Institute of Physics at Hebrew University where part of this work was completed.  This work is supported in part by the DOE grant DE-FG-02-92ER40704, and by an OJI award from the DOE.   

\appendix

\section{Feynman Integrals}
\label{app:ints}

For the calculation of order $\eta^1$ post-Minkowskian effects, the Feynman graphs in Fig.~\ref{fig:tail1} can be expressed as integrals of the form ($D=d-1$, with $d$ the spacetime dimension) 
\begin{equation}
\label{eq:unten}
I_{i_1\ldots i_n}  = \int \frac{d^D \mathbf q}{(2 \pi)^{D}} {{\bf q}_{i_1}\cdots {\bf q}_{i_n}\over \left( {\bf q}^2\right) \left[\big(\mathbf q + \mathbf k \right)^2 +\left(- \mathbf k^2 - i \epsilon\right) \big]},
\end{equation}
with $n\leq 4$.    Using rotational invariance, all such integrals can be converted (``tensor reduced") to linear combinations of tensors constructed from ${\bf k}^i$ and $\delta^{ij}$, with scalar integral coefficients of the form:
\begin{align} \label{eq_FMint2}
I_0(\alpha) & = \int \frac{d^{D} \mathbf q}{(2 \pi)^{D}} \frac{1}{\big[\mathbf q^2\big]^\alpha\left[\big(\mathbf q + \mathbf k \right)^2 +\left(- \mathbf k^2 - i \epsilon\right) \big]} \notag \\
                         & = \frac{1}{(4 \pi)^{D/2}} \frac{\Gamma(\alpha - D/2 + 1) \Gamma(D- 2 \alpha - 1)}{\Gamma(D-\alpha-1)} \left(- \mathbf k^2 - i \epsilon\right)^{D/2-\alpha-1}
\end{align}
which has been performed by standard textbook methods, see for instance~\cite{peskin}.    

At order $\eta^2$, the loop graphs in Fig.~\ref{fig:tail2}(a),(b) are expressible as nested integrals, which can be done by first evaluating a subgraph proportional to the standard integral
\begin{equation}
\label{eq_FMint1}
\int \frac{d^D \mathbf k}{(2 \pi)^D} \frac{1}{\left[(\mathbf k + \mathbf
p)^2 \right]^{n_1} \left[\mathbf k^2 \right]^{n_2}} = \frac{1}{(4
\pi)^{D/2}} \frac{\Gamma (n_1 + n_2 - D/2)}{\Gamma (n_1)
\Gamma(n_2)} \frac{\Gamma(D/2-n_1)
\Gamma(D/2-n_2)}{\Gamma(D-n_1-n_2)} (\mathbf p^2)^{D/2-n_1-n_2}
\end{equation}
The result of this integration then yields linear combinations of terms of the form Eq.~(\ref{eq:unten}), which can be evaluated using Eq.~(\ref{eq_FMint2}).    

The graph in Fig.~\ref{fig:tail2}(c) is more challenging.   Unlike Fig.~\ref{fig:tail2}(a),(b), it cannot be expressed in terms of simple nested one-loop integrals.   After tensor reductions, all terms in Fig.~\ref{fig:tail2}(c) can be expressed in terms of the scalar integrals ($n_{i=1,2,3} \in \mathbb Z$):
\begin{align}
\label{eq:I1}
I_1(\{n_i\})({\bf k}) & = \int \frac{d^D \mathbf l}{(2 \pi)^D} \frac{1}{\big[\left(\mathbf l - \mathbf k\right)^2\big]^{n_1} \, \big[\mathbf l ^2\big]^{n_2}} \int \frac{d^D \mathbf p}{(2 \pi)^D} \frac{1}{\big[\mathbf p^2\big]^{n_3} \, \big[\left(\mathbf p + \mathbf l \right)^2 +\left(- \mathbf k^2 - i \epsilon\right) \big]} \\
\label{eq:I2}
I_2(n_1, n_3) & = \int \frac{d^D \mathbf l}{(2 \pi)^D} \frac{1}{\big[\left(\mathbf l - \mathbf k\right)^2\big]^{n_1} \, \big[\mathbf l^2 + \left(- \mathbf k^2 - i \epsilon\right)\big]}\int \frac{d^D \mathbf p}{(2 \pi)^D} \frac{1}{\big[\mathbf p^2\big]^{n_3} \, \big[\left(\mathbf p + \mathbf l \right)^2 +\left(- \mathbf k^2 - i \epsilon\right) \big]} 
\end{align}
The common integral over $\mathbf p$ can be calculated by standard Feynman parameter methods and expressed as a hypergeometric function.
\begin{eqnarray}
\nonumber
I_{in}(n_3) & =& \int \frac{d^D \mathbf p}{(2 \pi)^D} \frac{1}{\big[\mathbf p^2\big]^{n_3} \, \big[\left(\mathbf p + \mathbf l \right)^2 +\left(- \mathbf k^2 - i \epsilon\right) \big]} \\
\nonumber
& = & \frac{1}{(4 \pi)^{D/2}} \frac{\Gamma(n_3 - D/2 + 1) \Gamma(D/2 -n_3)}{\Gamma(D/2)} \left(- \mathbf k^2 - i \epsilon\right)^{D/2-n_3-1}\\
\label{eq:hyperrep1}
& & {}\times  {}_2F_1 \left(n_3-D/2+1, n_3; D/2; \frac{\mathbf l^2}{\mathbf k^2 + i \epsilon}\right)
\end{eqnarray}	
Alternatively (and equivalently, using a well known integral representation for the hypergeometric function), one can introduce a Mellin-Barnes representation
\begin{equation}
{1\over \left(\mathbf p + \mathbf l \right)^2 +\left(- \mathbf k^2 - i \epsilon\right)} = \int_{-i\infty}^{i\infty} {dz\over 2\pi i} \Gamma(1+z) \Gamma(-z) {\left[({\bf p}+{\bf l})^2\right]^z\over \left(-{\bf k}^2-i\epsilon\right)^{z+1}},
\end{equation}
where the integration contour runs parallel to the $\mbox{Im} z$ axis in the way explained in~\cite{smirnov}.   In this representation
\begin{eqnarray}
\nonumber
I_{in}(n_3) & =& \frac{1}{(4 \pi)^{D/2}} \frac{\Gamma(D/2 - n_3) }{\Gamma(n_3)} \left(- \mathbf k^2 - i \epsilon\right)^{D/2-n_3-1}\\
\label{eq:MBin}
& & {}\times   \int_{-i\infty}^{+i\infty} \frac{dz}{2 \pi i} \frac{\Gamma(n_3 - D/2 + 1 + z) \Gamma(n_3 + z) \Gamma(-z)}{\Gamma(D/2+z)} \left(\frac{\mathbf l^2}{- \mathbf k^2 - i \epsilon}\right)^z.  
\end{eqnarray}	
Using Eq.~(\ref{eq:MBin}) together with Eq.~(\ref{eq_FMint1}), it is now possible to perform all momentum integrals in Eq.~(\ref{eq:I1}), with the result
\begin{eqnarray}
\nonumber
I_1(\{n_i\}) &=& {1\over (4\pi)^D}{\Gamma(D/2-n_1)\over \Gamma(n_1)} {\Gamma(D/2-n_3)\over \Gamma(n_3)}  \left(-{\bf k}^2-i\epsilon\right)^{D/2-n_3-1} \left({\bf k^2}\right)^{D/2-n_1-n_2} \\
\nonumber
& & {} \times \int_{-i\infty}^{i\infty} {dz\over 2\pi i} \left[{\Gamma(n_3+z) \Gamma(1+n_3-D/2+z)\Gamma(D/2-n_2+z) \over \Gamma(D/2+z) \Gamma(D-n_1-n_2+z)}\right.\\
& & {}\left. \,\,\,\,\,\,\,\,\,\,\,\,\,\,\,\,\,\,\,\,\,\,\,\,\,\,\,\,\,\,\,\,\, {\Gamma(n_1+n_2-z-D/2) \Gamma(-z)\over \Gamma(n_2-z)}  \left(-{1\over 1+i\epsilon}\right)^z\right].
\end{eqnarray}
This integral can now be evaluated by residues.   The result can be expressed in closed form as a sum over two generalized hypergeometric functions ${}_4 F_3$ evaluated at $z=1$.    It turns out that for the values of $n_i$ generated by the graph in Fig.~\ref{fig:tail2}(c), all such hypergeometric functions reduce to simple ratios of Gamma functions.    As the expressions are best handled by computer algebra, we will not reproduce them here.

For the second integral $I_2(n_1, n_3)$, we use a different contour integral representation of the hypergeometric function in Eq.~(\ref{eq:hyperrep1}) (see Eq. D.72 of~\cite{smirnov})
\begin{eqnarray}
\nonumber
I_{in}(n_3) & =& \frac{1}{(4 \pi)^{D/2}}{1\over \Gamma(n_3) \Gamma(D-1-n_3)} \int_{-i\infty}^{+i\infty} \frac{dz}{2 \pi i}\Big[\Gamma(1+n_3-D/2+z) \Gamma(n_3+z)\\
& & \times \Gamma(D-2 n_3 -1-z)\Gamma(-z) \left( {{\bf l}^2+(-{\bf k}^2-i\epsilon)\over - {\bf k}^2 - i \epsilon} \right)^z\Big].  
\end{eqnarray}	
Inserting this into Eq.~(\ref{eq:I2}), and performing the momentum integrals with the help of 
\begin{align}
J_0(\alpha, \beta) & = \int \frac{d^D \mathbf l}{(2 \pi)^D} \frac{1}{\big[\mathbf l^2\big]^\alpha\left[\big(\mathbf l + \mathbf k \right)^2 +\left(- \mathbf k^2 - i \epsilon\right) \big]^\beta} \notag \\ 
                  & = \frac{1}{(4 \pi)^{D/2}} \frac{\Gamma(\alpha + \beta - D/2) \Gamma(D - 2 \alpha - \beta)}{\Gamma(\beta) \Gamma(D-\alpha-\beta)} \left(- \mathbf k^2 - i \epsilon\right)^{D/2-\alpha-\beta},
\end{align}
we find the following contour integral representation for $I_2(n_2,n_3)$:
\begin{eqnarray}
\nonumber
I_2(n_2,n_3) &=& \frac{1}{(4 \pi)^{D}}\left(-{\bf k}^2-i\epsilon\right)^{D-n_1-n_3-2} {1\over \Gamma(n_3) \Gamma(D-1-n_3)}\\
\nonumber
& & \times  \int_{-i\infty}^{+i\infty} \frac{dz}{2 \pi i} \left[{\Gamma(1+n_3-D/2+z) \Gamma(n_3+z)\Gamma(D-2 n_1-1+z)\over \Gamma(D-n_1-1+z)}\right.\\
& &\left.  \,\,\,\,\,\,\,\,\,\,\,\,\,\,\,\,\,\,\,\,\,\,\,\,\,\,\,\,\, \,\,\,\,\,\,\,{\Gamma(D-2 n_3 -1-z) \Gamma(1+n_1-D/2-z) \Gamma(-z)\over\Gamma(1-z)} \right].
\end{eqnarray}
For the case $n_3=1$, the contour integral can be done using Barnes' second lemma (see~\cite{smirnov},  Eq. D.47), with the result
\begin{eqnarray}
\nonumber
I_2(n_1, 1) & =&  \frac{1}{D-3} \frac{\left(- \mathbf k^2 - i \epsilon\right)^{D-3-n_1}}{(4 \pi)^D} \left. \frac{\Gamma(n_1 - D/2 + 1)}{\Gamma(D-n_1-1)}\right[\Gamma(D - 2 n_1 - 1) \Gamma(2 - D/2)\\
& & \left.  {} - \frac{\Gamma(n_1-D+3) \Gamma(2 D - 2 n_1 - 4) \Gamma(D/2 - n_1) \Gamma(D/2 - 1)}{\Gamma(3D/2 - n_1 - 3) \Gamma(n_1)} \right].
\end{eqnarray}
For the case $n_3<1$, the contour integral is performed by taking residues.   It turns out that only the pole at $z=0$ contributes, and the result is
\begin{align}
I_2(n_1, n_3<1) & = \frac{\left(-\mathbf k^2 - i \epsilon\right)^{D-n_1-n_3-2}}{(4 \pi)^D}  \notag \\
& \times \frac{\Gamma(D-2n_1-1) \Gamma(D-2n_3-1) \Gamma(n_1-D/2+1) \Gamma(n_3-D/2+1)}{\Gamma(D-n_1-1) \Gamma(D-n_3-1)}.
\end{align}
We did not bother to evaluate $I_2(n_2,n_3>1)$, as such terms are not needed to compute Fig.~\ref{fig:tail2}(c).

\section{UV poles}
\label{app:poles}

After tensor reduction the amplitude for Fig.~\ref{fig:tail2}(c) is a linear combination of scalar integrals of the form
\begin{align}
I_{123} = \int_{\mathbf l, \mathbf p} \frac{1}{\big[\left(\mathbf l - \mathbf k\right)^2\big]^{n_1} \, \big[\mathbf l ^2\big]^{n_2} \, \big[\mathbf l^2 + \left(- \mathbf k^2 - i \epsilon\right)\big]} \frac{1}{\big[\mathbf p^2\big]^{n_3} \, \big[\left(\mathbf p + \mathbf l \right)^2 +\left(- \mathbf k^2 - i \epsilon\right) \big]} 
\end{align}
with $n_1, n_2, n_3 \in \mathbb Z$.   Here we have defined $\int_{\bf p}=\int d^D {\bf p}/(2\pi)^D$.   As in appendix~\ref{app:ints}, $D=d-1$.   In order to extract the UV behavior of the integrals we formally expand the denominator in powers of $- \mathbf k^2 - i \epsilon$, which yields
\begin{align}
I_{123} = \sum^\infty_{i,j=0} (-1)^{i+j} (- \mathbf k^2 - i \epsilon)^{i+j} \int_{\mathbf l, \mathbf p} \frac{1}{\big[\left(\mathbf l - \mathbf k\right)^2\big]^{n_1} \, \big[\mathbf l ^2\big]^{n_2+i+1} \, \big[\mathbf p ^2\big]^{n_3} \, \big[\left(\mathbf p + \mathbf l\right)^2\big]^{j+1}} \, . \label{eq:IintUVexpand}
\end{align}
The $\mathbf p$-integration is then easily performed using Eq. (\ref{eq_FMint1})
\begin{align}
I_{123} = & \sum_{i,j} (\mathbf k^2)^{i+j} \frac{1}{(4 \pi)^{\frac{D}{2}}} \frac{\Gamma(n_3 + j + 1 - D/2)}{\Gamma(n_3) \Gamma(j+1)} \frac{\Gamma(D/2-n_3) \, \Gamma(D/2-j-1)}{\Gamma(D-n_3-j-1)} \notag \\
& {} \ \ \ \ \ \times  \int_{\mathbf l} \frac{1}{\big[\left(\mathbf l + \mathbf k\right)^2\big]^{n_1} \, \big[\mathbf l ^2\big]^{n_2+n_3+i+j+2-\frac{D}{2}} } \label{eq:appB3},
\end{align}
and we note that the first line of Eq. (\ref{eq:appB3}) is free of any divergences for $d = 4$.
The remaining task is therefore to extract the UV poles from the integral over $\mathbf l$ in Eq. (\ref{eq:appB3})
which is\footnote{Here we took into account that $n_1 \le 1$ for the integrals needed so that the first factor in the denominator never yields an IR divergence.}
\begin{align}
\text{UV divergent if }& {} \ \ \ \ n_1 + n_2 + n_3 + i + j + 3 \le d ,\label{eq:uvdivcond}\\
\text{IR divergent if }& {} \ \ \ \ \ \ \ \ \ \ \, \hspace*{0.6pt} n_2 + n_3 + i + j + 3 \ge d \label{eq:irdivcond} \ .
\end{align}
Only a finite number of combinations of $i$ and $j$ can yield UV divergences which truncates the sums in Eq. (\ref{eq:appB3}).
If $n_1 \le 0$ it is possible that the integral is both IR and UV divergent, and our computation involves the cases $n_1 = 1, 0, -1, -2, -3$.

In the case of $n_1 = 1$ the UV divergences never overlap with the IR divergences and we can extract the UV divergences using Eq. (\ref{eq_FMint1}) and keeping only the UV poles. For $n_1 \ge 0$ the integration in Eq. (\ref{eq:appB3}) is a scaleless integral which vanishes in dimensional regularization. When it is logarithmically divergent however the vanishing implies a cancelation of IR and UV divergences. In order to extract the UV poles, we only need to consider this case and use
\begin{equation}
\int_{\mathbf l} \frac{1}{\big[\mathbf l ^2\big]^{D/2}} \rightarrow - \frac{1}{4 \pi^2} \frac{1}{(d-4)_{UV}}.
\end{equation}
In this way we find for the UV divergent part of diagram Fig.~\ref{fig:tail2}(c)
\begin{equation}
\left. {{\cal A}^{(c)}_{\eta^2}\over {\cal A}_{\eta^0}} \right|_{UV}    =    \left (G_N m |{\bf k}|\right)^2 \left[- {({\bf k}^2 + i\epsilon)\over \pi \mu^2} e^{\gamma_E}\right]^{(d-4)} \times \left[-{1046\over 315} {1\over (d-4)_{UV}}\right].
\end{equation}


\begin{thebibliography}{99}

\bibitem{LIGO}
A.~Abramovici {\it et al.},
Science {\bf 256}, 325 (1992);
A.~Giazotto,
Nucl.\ Instrum.\ Meth.\  A {\bf 289}, 518 (1990).


\bibitem{LISA}
K.~Danzmann and A.~Rudiger,
Class.\ Quant.\ Grav.\  {\bf 20}, S1 (2003).



\bibitem{PNrev}
L.~Blanchet,
Living Rev.\ Rel.\  {\bf 9}, 4 (2006).



\bibitem{BHpert}
M.~Sasaki and H.~Tagoshi,
Living Rev.\ Rel.\  {\bf 6}, 6 (2003)
[arXiv:gr-qc/0306120].


\bibitem{NREFT}
W.~E.~Caswell and G.~P.~Lepage,
Phys.\ Lett.\  B {\bf 167}, 437 (1986);
G.~T.~Bodwin, E.~Braaten and G.~P.~Lepage,
Phys.\ Rev.\  D {\bf 51}, 1125 (1995)
[Erratum-ibid.\  D {\bf 55}, 5853 (1997)]
[arXiv:hep-ph/9407339];
M.~E.~Luke, A.~V.~Manohar and I.~Z.~Rothstein,
Phys.\ Rev.\  D {\bf 61}, 074025 (2000)
[arXiv:hep-ph/9910209].



\bibitem{GnR1}
W.~D.~Goldberger and I.~Z.~Rothstein,
Phys.\ Rev.\  D {\bf 73}, 104029 (2006)
[arXiv:hep-th/0409156].

\bibitem{GnR1.5}
For reviews, see W.~D.~Goldberger and I.~Z.~Rothstein,
Gen.\ Rel.\ Grav.\  {\bf 38}, 1537 (2006)
[Int.\ J.\ Mod.\ Phys.\  D {\bf 15}, 2293 (2006)]
[arXiv:hep-th/0605238];
W.~D.~Goldberger,
``Les Houches lectures on effective field theories and gravitational
radiation,''
arXiv:hep-ph/0701129.


\bibitem{GnR3}
W.~D.~Goldberger and I.~Z.~Rothstein,
Phys.\ Rev.\  D {\bf 73}, 104030 (2006)
[arXiv:hep-th/0511133].

\bibitem{Porto2}
R.~A.~Porto,
Phys.\ Rev.\  D {\bf 77}, 064026 (2008)
[arXiv:0710.5150 [hep-th]].

\bibitem{Porto1}
R.~A.~Porto,
Phys.\ Rev.\  D {\bf 73}, 104031 (2006)
[arXiv:gr-qc/0511061].

\bibitem{PR1}
R.~A.~Porto and I.~Z.~Rothstein,
Phys.\ Rev.\ Lett.\  {\bf 97}, 021101 (2006)
[arXiv:gr-qc/0604099];
R.~A.~Porto and I.~Z.~Rothstein,
arXiv:0712.2032 [gr-qc];
R.~A.~Porto and I.~Z.~Rothstein,
Phys.\ Rev.\  D {\bf 78}, 044012 (2008)
[arXiv:0802.0720 [gr-qc]].

\bibitem{PR2}
R.~A.~Porto and I.~Z.~Rothstein,
Phys.\ Rev.\  D {\bf 78}, 044013 (2008)
[arXiv:0804.0260 [gr-qc]].

\bibitem{galley1}
C.~R.~Galley and B.~L.~Hu,
arXiv:0801.0900 [gr-qc].

\bibitem{galley2}
C.~R.~Galley and M.~Tiglio,
arXiv:0903.1122 [gr-qc].

\bibitem{Mino:1996nk}
Y.~Mino, M.~Sasaki and T.~Tanaka,
Phys.\ Rev.\  D {\bf 55}, 3457 (1997)
[arXiv:gr-qc/9606018].

\bibitem{Quinn:1996am}
T.~C.~Quinn and R.~M.~Wald,
Phys.\ Rev.\  D {\bf 56}, 3381 (1997)
[arXiv:gr-qc/9610053].

 \bibitem{Kol:2007rx}
B.~Kol and M.~Smolkin,
Phys.\ Rev.\  D {\bf 77}, 064033 (2008)
[arXiv:0712.2822 [hep-th]].

\bibitem{Kol:2007bc}
B.~Kol and M.~Smolkin,
Class.\ Quant.\ Grav.\  {\bf 25}, 145011 (2008)
[arXiv:0712.4116 [hep-th]].

\bibitem{Kol:2009mj}
B.~Kol and M.~Smolkin,
arXiv:0910.5222 [hep-th].


\bibitem{Gilmore}
J.~B.~Gilmore and A.~Ross,
Phys.\ Rev.\  D {\bf 78}, 124021 (2008)
[arXiv:0810.1328 [gr-qc]].

\bibitem{Cannella:2008nr}
  U.~Cannella and R.~Sturani,
  arXiv:0808.4034 [gr-qc].

\bibitem{Cannella:2009he}
  U.~Cannella, S.~Foffa, M.~Maggiore, H.~Sanctuary and R.~Sturani,
  arXiv:0907.2186 [gr-qc].

\bibitem{KKBHrefs}
B.~Kol,
Phys.\ Rept.\  {\bf 422}, 119 (2006)
[arXiv:hep-th/0411240];
 T.~Harmark and N.~A.~Obers,
  ``Phases of Kaluza-Klein black holes: A brief review,''
  arXiv:hep-th/0503020.

\bibitem{GnR2.5}
Y.~Z.~Chu, W.~D.~Goldberger and I.~Z.~Rothstein,
JHEP {\bf 0603}, 013 (2006)
[arXiv:hep-th/0602016].

\bibitem{smolkin}
J.~B.~Gilmore, A.~Ross and M.~Smolkin,
JHEP {\bf 0909}, 104 (2009)
[arXiv:0908.3490 [hep-th]].

\bibitem{thorne}
For a review of different multipole moment formalisms, see K.~S.~Thorne,
Rev.\ Mod.\ Phys.\  {\bf 52}, 299 (1980).

\bibitem{maggiore}
M.~Maggiore, ``Gravitational Waves. Vol. 1: Theory and Experiments,''
{\it  Oxford University Press, October 2007. 572p. (ISBN-13: 978-0-19-857074-5)}


\bibitem{burke}
W. L. Burke, J. Math. Phys. {\bf 12} 401 (1971).

\bibitem{postM1}
L.~Blanchet and T.~Damour,
  Phil.\ Trans.\ Roy.\ Soc.\ Lond.\  A {\bf 320}, 379 (1986).

\bibitem{tail1}
  L.~Blanchet and T.~Damour,
  Phys.\ Rev.\  D {\bf 37}, 1410 (1988);
L.~Blanchet and T.~Damour,
  Phys.\ Rev.\  D {\bf 46}, 4304 (1992).

 

\bibitem{postM2}  
M.~E.~Pati and C.~M.~Will,
  Phys.\ Rev.\  D {\bf 62}, 124015 (2000)
  [arXiv:gr-qc/0007087];
M.~E.~Pati and C.~M.~Will,
  Phys.\ Rev.\  D {\bf 65}, 104008 (2002)
  [arXiv:gr-qc/0201001];
C.~M.~Will,
  Phys.\ Rev.\  D {\bf 71}, 084027 (2005)
  [arXiv:gr-qc/0502039].

 

\bibitem{tail}
E.~T.~Newman and R.~Penrose,
  Proc.\ Roy.\ Soc.\ Lond.\  A {\bf 305}, 175 (1968);
  J.~M.~Bardeen and W.~H.~Press,
  J.\ Math.\ Phys.\  {\bf 14}, 7 (1973).
  
 

  
 \bibitem{Blanchet:1997jj}
L.~Blanchet,
Class.\ Quant.\ Grav.\  {\bf 15}, 113 (1998)
[Erratum-ibid.\  {\bf 22}, 3381 (2005)]
[arXiv:gr-qc/9710038].

\bibitem{Weinberg:1965nx}
S.~Weinberg,
Phys.\ Rev.\  {\bf 140}, B516 (1965).

 \bibitem{smirnov}
  V.~A.~Smirnov,
  ``Feynman integral calculus,''
{\it  Berlin, Germany: Springer (2006)}.


\bibitem{Anderson:1982fk}
J.~L.~Anderson, L.~S.~Kegeles, R.~G.~Madonna and R.~E.~Kates,
Phys.\ Rev.\  D {\bf 25}, 2038 (1982).









\bibitem{IraEFT}
I.~Z.~Rothstein, ``TASI lectures on effective field theories,''  arXiv:hep-ph/0308266.


\bibitem{approximants}
 See for instance, T.~Damour, B.~R.~Iyer and B.~S.~Sathyaprakash,
  Phys.\ Rev.\  D {\bf 57}, 885 (1998)
  [arXiv:gr-qc/9708034];
  A.~Buonanno and T.~Damour,
  Phys.\ Rev.\  D {\bf 59}, 084006 (1999)
  [arXiv:gr-qc/9811091].






\bibitem{Khriplovich:1997ms}
I.~B.~Khriplovich and A.~A.~Pomeransky,
Phys.\ Lett.\  A {\bf 252}, 17 (1999)
[arXiv:gr-qc/9712040].

\bibitem{Asada:1997zu}
 H.~Asada and T.~Futamase,
 Phys.\ Rev.\  D {\bf 56}, 6062 (1997)
 [arXiv:gr-qc/9711009].
 



\bibitem{Damour:2007xr}
  T.~Damour and A.~Nagar,
  Phys.\ Rev.\  D {\bf 76}, 064028 (2007)
  [arXiv:0705.2519 [gr-qc]].
 
 
 \bibitem{PRR}
  R.~A.~Porto, A.~Ross and I.~Z.~Rothstein,
  arXiv:1007.1312 [gr-qc].


 \bibitem{Blanchet:1997ji}
  L.~Blanchet,
  Class.\ Quant.\ Grav.\  {\bf 15}, 89 (1998)
  [arXiv:gr-qc/9710037].


\bibitem{memory}
D.~Christodoulou,
  Phys.\ Rev.\ Lett.\  {\bf 67}, 1486 (1991).
  

\bibitem{Wiseman:1991ss}
  A.~G.~Wiseman and C.~M.~Will,
  Phys.\ Rev.\  D {\bf 44}, 2945 (1991).

  
  
\bibitem{peskin}
  M.~E.~Peskin and D.~V.~Schroeder,
  ``An Introduction To Quantum Field Theory,''
  {\it  Reading, USA: Addison-Wesley (1995)}.






\end{thebibliography}
\end{document}